\documentclass[twoside]{article}
\usepackage[latin1]{inputenc}
\usepackage{t1enc}
\usepackage{a4}
\usepackage{tabularx}
\usepackage{epsf}
\input psfig.tex

\def\bref{\vspace{4pt}\noindent\hangindent=10mm}

\begin{document}

\setcounter{figure}{0}
\setcounter{section}{0}
\setcounter{equation}{0}

\begin{center}
{\Large\bf New Cosmology with Clusters of
Galaxies}\,\,\\[0.7cm]

Peter Schuecker \\[0.17cm]
Max-Planck-Institut f\"ur extraterrestrische Physik \\
Giessenbachstra{\ss}e, Postfach 1312, D-85741 Garching \\
peters@mpe.mpg.de
\end{center}

\vspace{0.5cm}

\begin{abstract}
\noindent{\it 
The review summarizes present and future applications of galaxy
clusters to cosmology with emphasis on nearby X-ray clusters. The
discussion includes the density of dark matter, the normalization of
the matter power spectrum, neutrino masses, and especially the
equation of state of the dark energy, the interaction between dark
energy and ordinary matter, gravitational holography, and the effects
of extra-dimensions.}
\end{abstract}

\section{Basic cosmological framework}\label{BASIC}

The general framework for present cosmological work is set by three
observational results. The perfect Planckian shape of the cosmic
microwave background (CMB) spectrum as observed with the COBE
satellite (Mather et al. 1990) clearly shows that the Universe must
have evolved -- from a hot, dense, and opaque phase. The very good
correspondence of the observed abundance of light elements and the
results of Big Bang Nucleosynthesis (BBN, e.g. Burles, Nollett \&
Turner 2001) shows that the cosmic expansion can be traced back to
cosmological redshifts up to $z=10^{10}$. Steigman (2002) pointed out
that if these analyses would have been performed with Newton gravity
and not with Einstein gravity, then the observed abundances could not
be reconciled with the BBN predictions. One can take this as one of
the few indications than Einstein gravity can in fact be applied
within a cosmological context and underlines the importance of the BBN
benchmark for any gravitational theory. Finally, the consistency of
the ages of the oldest stars in globular clusters (e.g. Chaboyer \&
Krauss 2002) and the age of the Universe as obtained from cosmological
observations can be regarded as the long-waited `unification' of the
theory of stellar structure and the theory of cosmic spacetime
(Peebles \& Ratra 2004). Traditionally, Friedmann-Lema\^{\i}tre (FL)
world models as derived from Einstein's field equations for spatially
homogeneous and isotropic systems, are assumed, characterized by the
Hubble constant $H_0$ in units of $h= H_0/(100\,{\rm km}\,{\rm
s}^{-1}\,{\rm Mpc}^{-1})$, the normalized density of cosmic matter
$\Omega_{\rm m}$ (e.g., baryonic and Cold Dark Matter CDM), the
normalized cosmological constant $\Omega_\Lambda$, and its equation of
state $w$.  Within this general framework, clusters of galaxies are
traditionally used as cosmological probes on Gigaparsec scales.
However, a precise test that one can apply Einstein gravity on such
large scales is still missing.

In Sect.\,\ref{GCLST}, a summary of the basic properties of nearby
galaxy clusters is given. The hierarchical structure formation
paradigm is tested with nearby galaxy clusters in
Sect.\,\ref{HIER}. Constraints on the density of dark matter (DM), the
normalization of the matter power spectrum, and neutrino masses are
presented in Sect.\,\ref{MATTER}. Observational effects of the
equation of state of the dark energy (DE), and a first test of a
non-gravitational interaction between DE and DM are presented in
Sect.\,\ref{DE}. The problem of the cosmological constant and its
discussion in terms of the gravitional Holographic Principle as well
as the effect of an extra-dimension of brane-world gravity are
discussed in Sect.\,\ref{CCP}. Sect.\,\ref{FUTURE} draws some
conclusions. A general review on clusters is given in Bahcall (1999),
whereas Edge (2004) focuses on nearby X-ray cluster surveys, Borgani
\& Guzzo (2001) on their spatial distribution, Rosati, Borgani \&
Norman (2002) and Voit (2004) on their evolution.

\section{Galaxy clusters}\label{GCLST}

Galaxy clusters are the largest virialized structures in the
Universe. Only 5\% of the bright galaxies ($>L_*$) are found in rich
clusters, but more than 50\% in groups and poor clusters. The number
of cluster galaxies brighter than $m_3+2^m$ where $m_3$ is the
magnitude of the third-brightest cluster galaxy, and located within
$1.5\,h^{-1}\,{\rm Mpc}$ radius from the cluster center, range for
rich clusters from 30 to 300 galaxies, and for groups and poor
clusters from 3 to 30. For cosmological tests, rich clusters will turn
out to be of more importance so that the following considerations will
mainly focus on the properties of this type. Rich clusters have
typical radii of $1-2\,h^{-1}\,{\rm Mpc}$ where the surface galaxy
density drops to $\sim 1$\% of the central density.

Baryonic gas, falling into the cluster potential well, is shock-heated
up to temperatures of $T_{\rm e}=10^{7-8}\,{\rm K}$. The acceleration
of the electrons in the hot plasma (intracluster medium ICM) gives
thermal Bremsstrahlung with a maximum emissivity at $k_{\rm B}T_{\rm
e}\,=\,2-14\,{\rm keV}$ so that they can be observed in X-rays
together with some line emission. Typical X-ray luminosities range
between $L_{\rm x}=10^{42-45}\,h^{-2}\,{\rm erg}\,{\rm s}^{-1}$ in the
energy interval $0.1-2.4$\,keV. With X-ray satellites like ROSAT,
Chandra, or XMM-Newton, these clusters can thus be detected up to
cosmological interesting redshifts. However, only a few clusters are
detected at redshifts beyond $z=1$ (Rosati et al. 2002). 

Galaxy clusters are rare objects with number densities of
$10^{-5}\,h^3$ ${\rm Mpc}^{-3}$, strongly decreasing with X-ray
luminosity or cluster mass (B\"ohringer et al. 2002).  Current
structure formation models predict of the order of $10^6$ rich galaxy
clusters in the visible Universe, the majority with redshifts below
$z=2$. More than 5\,000 nearby galaxy clusters are already identified
in the optical as local concentrations of galaxies, and 2\,000 by
their (extended) X-ray emission. Surveys planned for the next few
years like the Dark Universe Observatory DUO (Griffiths, Petri,
Hasinger et al. 2004) will yield about $10^4$ clusters possibly up to
$z=2$, that is, already 1\% of the total cluster population. It
appears thus not completely illusory to finally get an almost complete
census of all rich galaxy clusters in the visible Universe.

X-ray clusters get their importance for cosmology because of the tide
correlations between observables like X-ray temperature or X-ray
luminosity and total gravitating cluster mass which allow a precise
reconstruction of the cosmic mass distribution on large scales.

Knowledge of the total gravitating mass of a cluster within a
well-defined radius, is of crucial importance. The masses are
summarized in cluster mass functions which depend on structure
formation models through certain values of the cosmological
parameters. However, cosmic mass function appear to be independent of
cosmology when they are written in terms of natural ``mass'' and
``time'' variables (Lacey \& Cole 1994).  Model mass functions can
either be predicted from semi-analytic models (e.g, Sheth \& Tormen
2002, Schuecker et al. 2001a, Amossov \& Schuecker 2004) or from
N-body simulations, the latter with errors between 10 to 30\% (Jenkins
et al. 2001).

Cluster masses can be determined in the optical by the velocity
dispersion of cluster galaxies or in X-rays from, e.g., the gas
temperature and density profiles, assuming virial and hydrostatic
equilibrium, respectively (and spherical symmetry). Gravitational
lensing uses the distortion of background galaxies and determines the
projected cluster mass without any specific assumption (e.g., Kaiser
\& Squires 1993).  For regular clusters, the masses of galaxy clusters
are consistently determined with the three methods and range between
$10^{14}-10^{15}\,h^{-1}\,M_\odot$ (e.g., Wu et al. 1998). Several
projects are currently under way to compare the mass estimates
obtained with the different methods in more detail. The baryonic mass
in clusters comes from the ICM and the stars in the cluster
galaxies. The ratio between the baryonic and total gravitating mass
(baryon fraction) in a cluster is about $0.07h^{-1.5}+0.05$.

Systematic X-ray studies of large samples of galaxy clusters have
revealed that about half of the clusters have significant substructure
in their surface brightness distributions, i.\,e., some deviations
from a perfect regular shape (e.g. Schuecker et al. 2001b). For the
detection of substructure, different methods as summarized in Feretti,
Giovannini \& Gioa (2002) give substructure occurence rates ranging
from 20 to 80\%. The large range clearly shows that the definition of
a well-defined mass threshold for substructure and the measurement of
the masses of the different subclumps is difficult and has not yet
been regorously applied. Further interesting ambiguities arise because
clusters appear more regular in X-ray pseudo pressure maps (product of
projected gas mass density and gas temperature) whereas contact
discontinuities and shock fronts caused by merging events appear more
pronounced in pseudo entropy and temperature maps (Briel, Finoguenov
\& Henry 2004).

Substructering is taken as a signature of the dynamical youth of a
galaxy cluster. The most dramatic distortions occure when two big
equal mass clumps collide (major merger) to form a larger
cluster. With the ROSAT satellite, merging events could be studied for
the first time in X-rays in more detail (e.g. Briel, Henry \&
B\"ohringer 1992). A typical time scale of a merger event is
$10^9$\,yr where the increased gas density and X-ray temperature can
boost X-ray luminosities up to factors of five (Randall, Sarazin \&
Ricker 2002). The XMM-Newton and especially the Chandra X-ray
satellite allows more detailed studies of substructures down to arcsec
scales. Substructures in form of cavities and bubbles (B\"ohringer et
al. 1993, Fabian et al. 2000), cold fronts (Vikhlinin, Markevitch \&
Murray 2001), weak shocks and sound waves (Fabian et al. 2003), strong
shocks (Forman et al. 2003), and turbulence (Schuecker et al. 2004)
were discovered, possibly triggered by merging events and/or AGN
activity. With the ASTRO-E2 satellite planned to be launched in 2005,
the line-of-sight kinematics of the ICM will be studied for the first
time to get more information about the dynamical state of the ICM. The
majority of the abovementioned substructures have low amplitudes which
do not much disturb radially-averaged cluster profiles (after masking)
and thus cluster mass estimates. In fact, the hydrostatic equation
relates the observed smooth pressure gradients to the total
gravitating cluster mass, which makes the robustness of X-ray cluster
mass estimates from numerical simulations plausible (Evrard, Metzler
\& Navarro 1996, but see Sect.\,\ref{FUTURE}). Present cosmological
tests based on galaxy clusters assume that the diversity of regular
and substructured clusters contribute only to the intrinsic scatter of
the observed X-ray luminosity-mass relation or similar diagnostics,
while keeping the shape and normalization of the original relation
almost unaltered.

\begin{figure}[]
\vspace{0.0cm}
\center{\hspace{-5.5cm}\psfig{figure=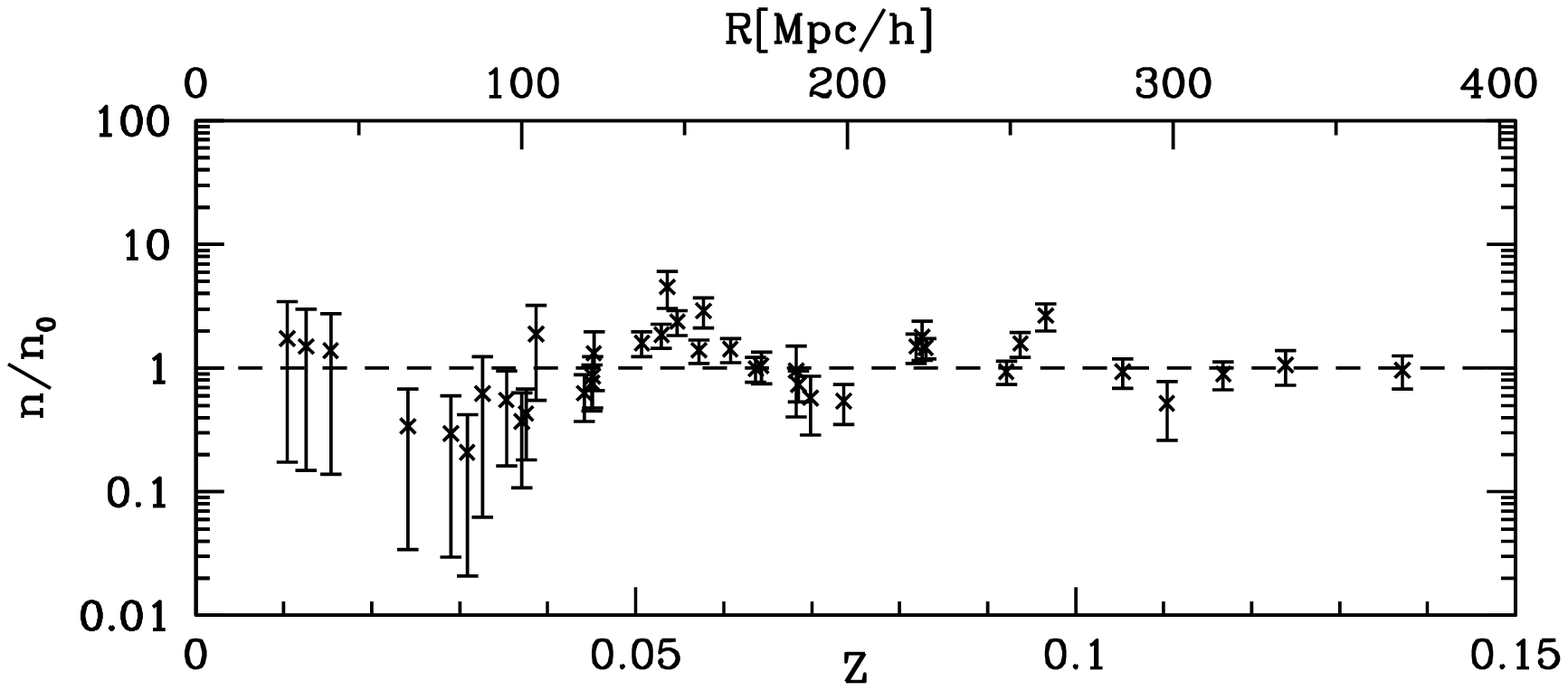,height=4.5cm}}
\vspace{-5.3cm}
\center{\hspace{+6.3cm}\psfig{figure=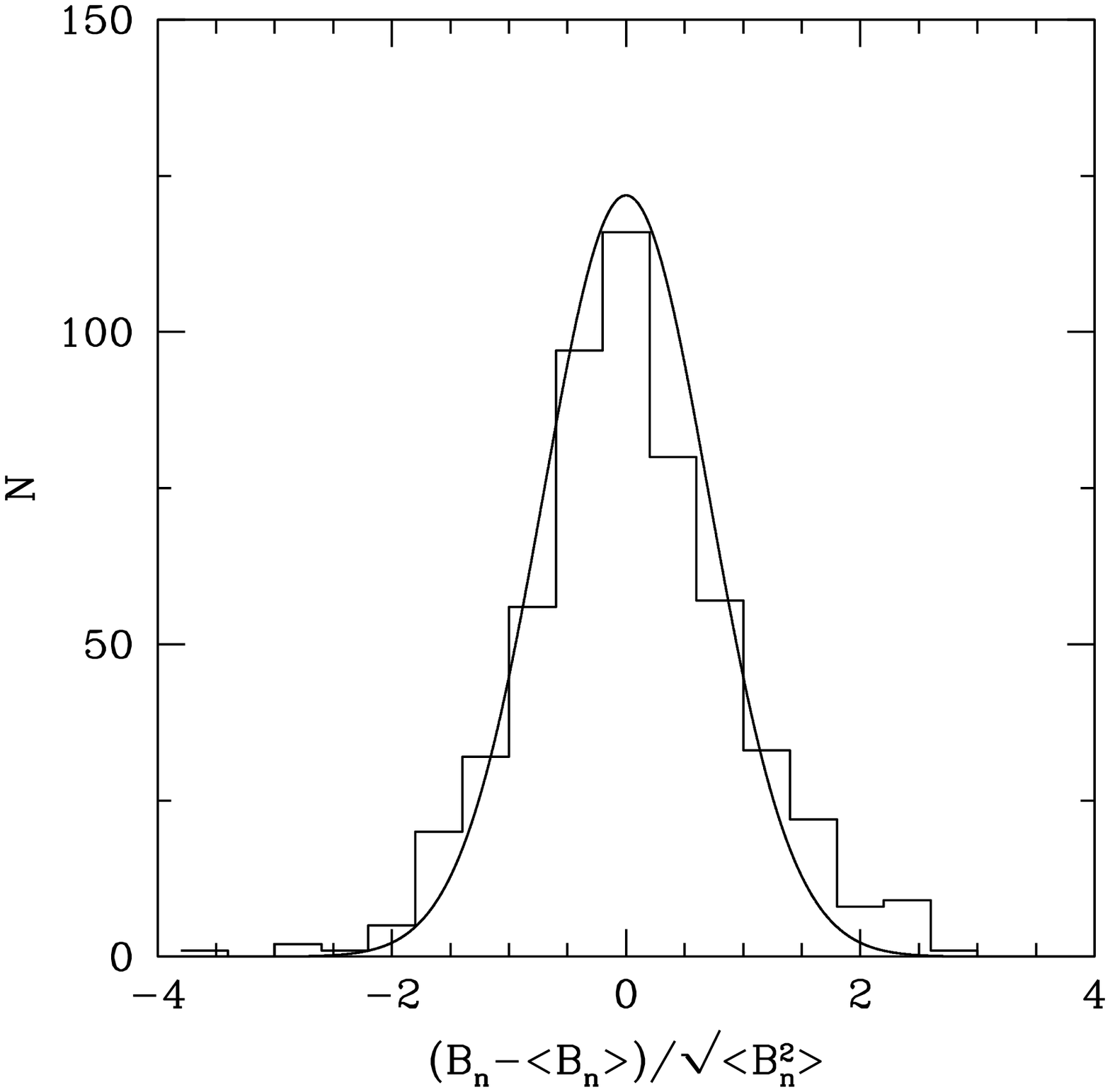,height=5.0cm}}
\vspace{-0.5cm}
\caption[]{\small {\bf Left:} Normalized comoving REFLEX cluster number
densities as a function of redshift, and comoving radial distance
$R$. Vertical error bars represent the formal $1\sigma$ Poisson
errors. {\bf Left:} Histogram of the normalized KL coefficients of the
REFLEX sample and superposed Gaussian profile. The Kolmogorov-Smirnov
probability for Gaussianity is 93\%.}
\label{FIG_NZ}
\end{figure}

The remaining about 50\% of the clusters appear quite regular - a
significant fraction of these clusters have very bright X-ray cores,
where the dense gas could significantly cool. Such cooling core
clusters are expected to be in a very relaxed dynamical state since
several Gigayears. Numerical simulations suggest that the baryon
fraction in these clusters is close to the universal value and can be
used after some corrections as a cosmic `standard candle' (e.g. White
et al. 1993).

For nearby ($z<0.3$) rich systems, evolutionary effects on core radius
and entropy input are found to be negligible (Rosati et
al. 2002). Detailed XMM studies at $z\sim 0.3$ can be found in Zhang
et al. (2004). Therefore, cosmological tests based on massive nearby
clusters with gas temperatures $k_{\rm B}T_{\rm e}>3$\,keV are
expected to give reliable results. For these systems, the observed
X-ray luminosity can be transformed into the theory-related cluster
mass with empirical luminosity-mass or similar relations characterized
by their shape, intrinsic scatter, and normalization (e.g., Reiprich
\& B\"ohringer 2002). It will be shown that with such methods, 
cosmological tests can be performed presently on the 20-30\% accuracy
level.

Further improvements on cluster scaling relations are thus necessary
to reach (if possible) the few-percent level of `precision
cosmology'. Large and systematic observational programms based on
Chandra and XMM-Newton observations are now under way which are
expected to significantly improve the relations within the next few
years (e.g., XMM Large Programme, H. B\"oh\-ringer et al., in prep.,
and a large Chandra project of T.H. Reiprich et al., in prep.). For
cosmological tests with distant rich clusters, additional work is
necessary. Gravi\-ta\-tio\-nal\-ly-induced evolutionary effects due to
structure growth, and non-gravitationally-induced evolutionary effects
like ICM heating through galactic winds caused by supernovae (SNe),
and heating by AGN cause systematic deviations from simple
self-similarity expectations (Kaiser 1986, Ponman, Cannon \& Navarro
1999). For cosmological tests, such evolutionary effects add further
degrees of freedom to be determined simultaneously with the
cosmological parameters (e.g., Borgani et al. 1999).

\section{Hierarchical structure formation paradigm}\label{HIER}

Structure formation on the largest scales as probed by galaxy clusters
is mainly driven by gravity and should thus be understandable in a
simple manner. However, reconciling the tiny CMB anisotropies at
$z\approx 1100$ with the very large inhomogeneities of the local
galaxy distribution has shown that the majority of cosmic matter
must come in nonbaryonic form (e.g., CDM). A direct consequence of
such scenarios is that clusters should grow from Gaussian initial
conditions in a quasi hierarchical manner, i.e., less rich clusters
and groups tend to form first and later merge to build more massive
clusters. The merging of galaxy clusters as seen in X-rays
(Sect.\,\ref{GCLST}) is a direct indication that such processes are
still at work in the local universe. 

A further argument for hierarchical structure growth comes from the
spatial distribution of galaxy clusters on $10^2\,h^{-1}\,{\rm Mpc}$
scales. Less then 1/10 of this distance can be covered by cluster
peculiar velocities within a Hubble time, keeping in this linear
regime the Gaussianity of the cosmic matter field as generated by the
chaotic processes in the early Universe almost intact. This
Gaussianity formally stems from the random-phase superposition of
plane waves and the central limit theorem (superposition
approximation). The peaks of this random field will eventually
collapse to form virialized clusters. The relation between the spatial
fluctuations of the clusters and the underlying matter field is called
`biasing'. For Gaussian random fields, the biasing tend to concentrate
the clusters in regions with the highest global matter density in a
manner that their correlation strengths $r_0$ increase with cluster
mass (Kaiser 1984) - otherwise they would immediately distroy
Gaussianity (e.g. if we would put a very massive cluster into a void
of galaxies). Peculiar velocities of the clusters induced by the
resulting inhomogeneities modify the $r_0$-mass relation, but without
disturbing the general trend (peak-background split of Efstathiou,
Frank \& White 1988, \& Mo \& White 1996). In the linear regime, we
thus expect a Gaussian distribution of the amplitudes of cluster
number fluctuations which increase with mass in a manner as predicted
by the specific hierarchical scenario.

\begin{figure}[]
\vspace{0.0cm}
\center{\hspace{-6.5cm}\psfig{figure=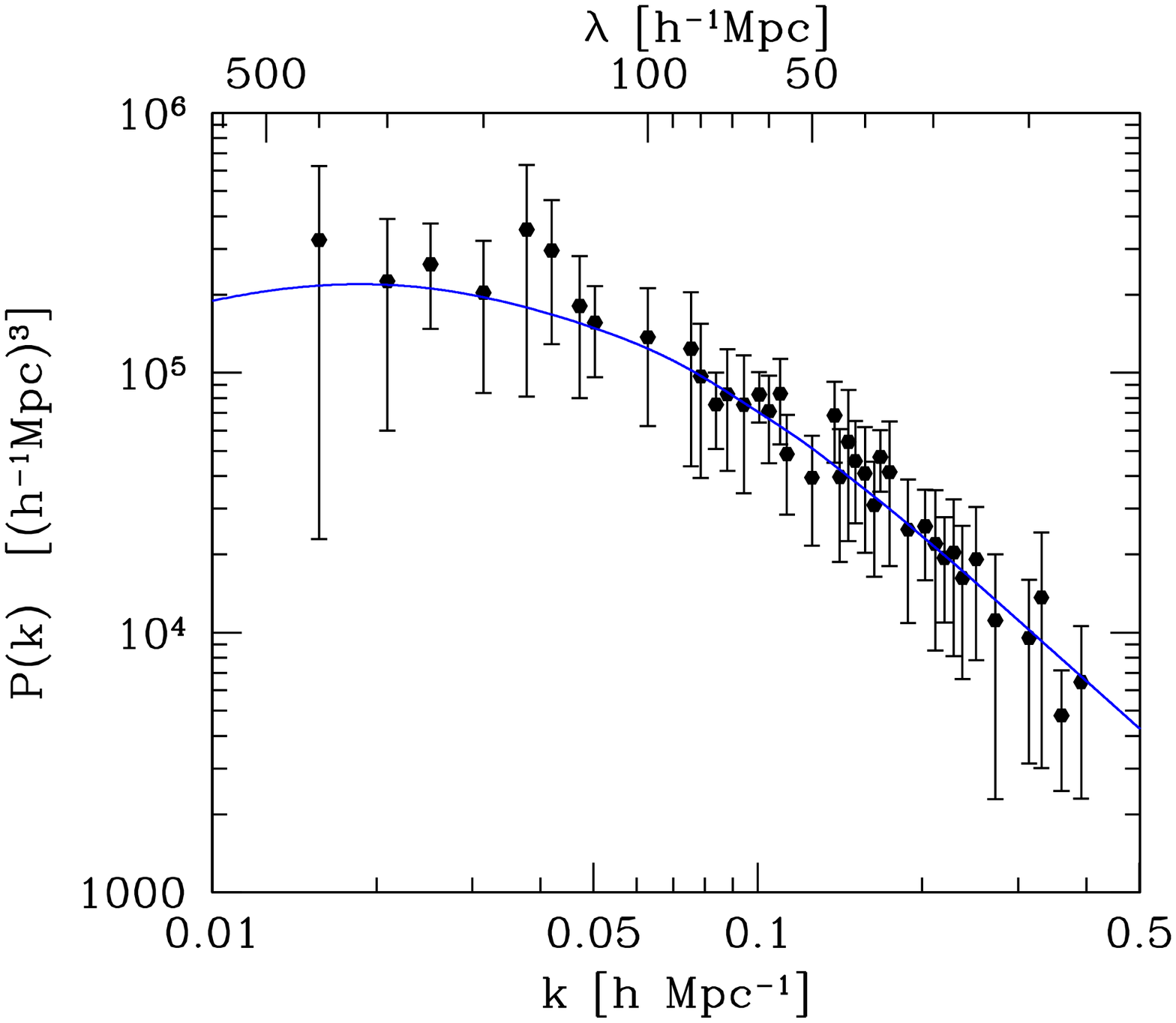,height=6.0cm}}
\vspace{-6.3cm}
\center{\hspace{+5.3cm}\psfig{figure=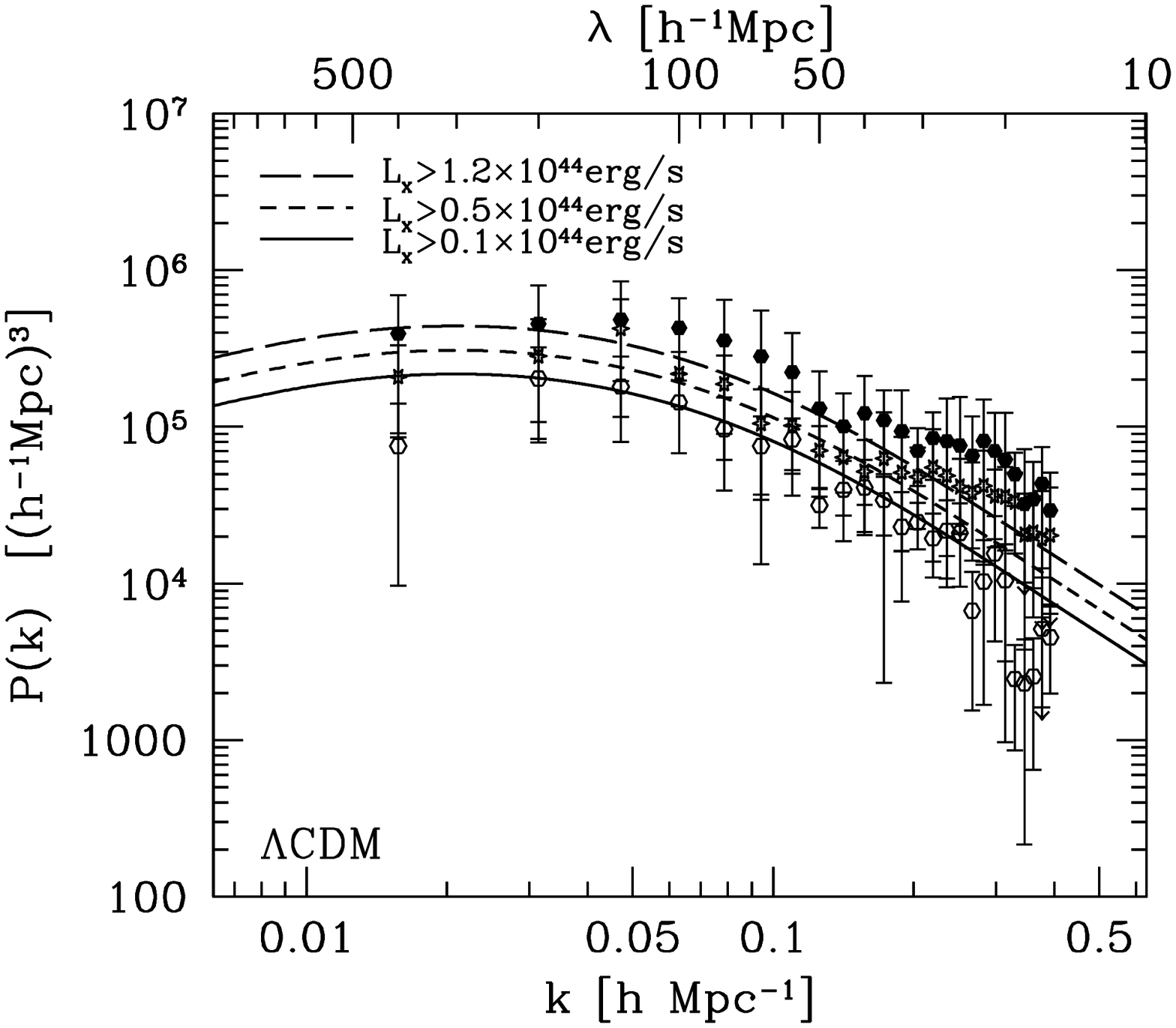,height=6.0cm}}
\vspace{-0.5cm}
\caption[]{\small {\bf Left:} Comparison of the observed REFLEX power
spectrum (points with error bars) with the prediction of a spatially
flat $\Lambda$CDM model with a matter density of $\Omega_{\rm m}=0.34$
and $\sigma_8=0.71$. Errors include cosmic variance and are estimated
with numerical simulations. {\bf Right:} Comparison of observed power
spectral densities and predictions of a low-density CDM semi-analytic
model as a function of lower X-ray luminosity, i.e., lower X-ray
cluster mass (Schuecker et al. 2001c). The errors include cosmic
variance and are obtained from N-body simulations.}
\label{FIG_DECO}
\end{figure}

The REFLEX catalogue (B\"ohringer et al. 2004)\footnote{
http://www.xray.mpe.mpg.de/theorie/REFLEX/} provides the largest
homogeneously selected sample of X-ray clusters and is ideally suited
for testing specific hierarchical structure formation models. The
sample comprises 447 southern clusters with redshifts $z\le0.45$
(median at $z=0.08$) down to X-ray fluxes of $3.0\times 10^{-12}\,{\rm
erg}\,{\rm s}^{-1}\,{\rm cm}^{-2}$ in the energy range
$(0.1-2.4)$\,keV, selected from the ROSAT All-Sky Survey (B\"ohringer
et al. 2001). Several tests show that the sample cannot be seriously
affected by unknown selection effects. An illustration is given by the
normalized, radially-averaged comoving number densities along the
redshift direction (Fig.\,\ref{FIG_NZ} left). The densities fluctuate
around a $z$-independent mean as expected when no unknown selection or
evolutionary effects are present. For further tests can be found in
B\"ohringer et al. (2001, 2004), Collins et al. (2000), and Schuecker
et al. (2001c).

Tests of the Gaussianity of the cosmic matter field refer to the
superposition approximation mentioned above. They devide the survey
volume into a set of large non-overlapping cells, count the clusters
in each cell, decompose the fluctuation field of the cluster counts
into plane waves via Fourier transformation, and check whether the
frequency distribution of the amplitudes of the plane waves (Fourier
modes with wavenumber $k$) follow a Gaussian distribution. However,
the survey volume provides only a truncated view of the cosmic matter
field which will result in an erroneous Fourier transform (the result
obtained will be the convolution of the true Fourier transform with
the survey window function). The truncation effect comprises both the
reduction of fine details in the Fourier transform and the correlation
of Fourier modes so that fluctuation power migrates between the
modes. This leakage effect increases when the symmetry of the survey
volume deviates from a perfect cubic shape. Uncorrelated amplitudes
can be obtained, when the fluctuations are decomposed into modes which
follow to some extent the shape of the survey volume. The
Karhunen-Lo{\`e}we (KL) decomposition determines such eigenmodes under
the constraint that the resulting KL fluctuation amplitudes are
statistically uncorrelated. This construction is quite optimal for
testing cosmic Gaussianity. The KL eigenmodes are the eigenvectors of
the sample correlation matrix, i.e., the matrix giving the expected
correlations between the number of clusters obtained in pairs of count
cells as obtained with a fiducial (e.g. concordance) cosmological
model. KL modes were first applied to CMB data by Bond (1995), to
galaxy data by Vogeley \& Szalay (1996), and to cluster data by
Schuecker et al. (2002). The linearity of the KL transform and the
direct biasing scheme expected for galaxy clusters suggest that the
statistics of the KL coefficients should directly reflect the
statistics of the underlying cosmic matter field.

Figure\,\ref{FIG_NZ} (right) compares a standard Gaussian with the
frequency distribution of the observed KL-transformed and normalized
cluster counts obtained with REFLEX. The cell sizes are larger than
$100\,h^{-1}\,{\rm Mpc}$ and thus probe Gaussianity in the linear
regime. The observed Gaussianity of the REFLEX data suggests
Gaussianity of the underlying cosmic matter field on such large
scales. This is a remarkable finding, taking into account the
difficulties one has to test Gaussianity even with current CMB data
(Komatsu et al. 2003, Cruz et al. 2004).

\begin{figure}[h]
\vspace{-0.0cm}
\center{\hspace{-1.0cm}\psfig{figure=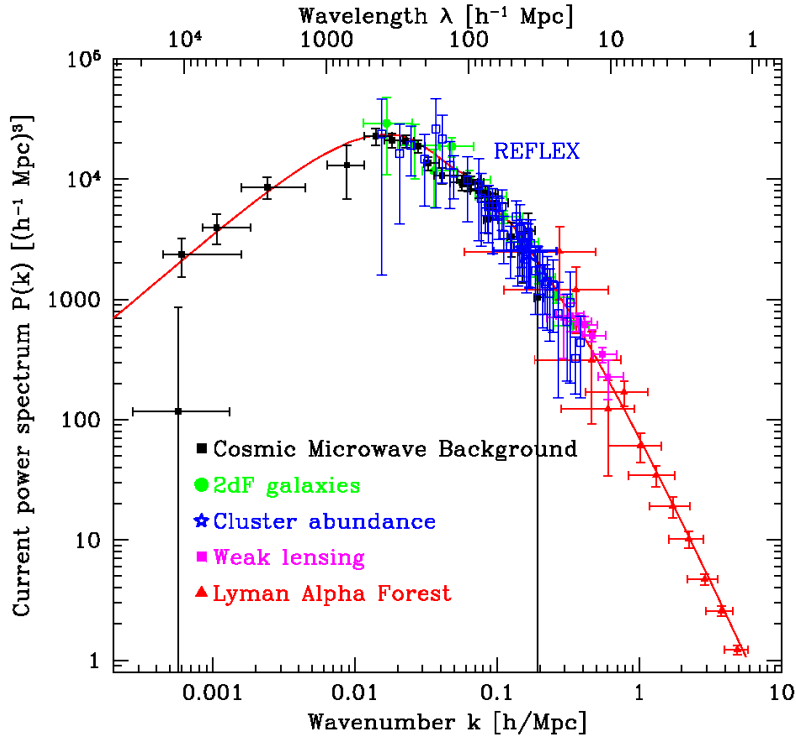,height=12.0cm}}
\vspace{-1.5cm}
\caption[]{\small Compilation of fluctuation power spectra of various
cosmological objects as complited by Tegmark \& Zaldarriaga (2002)
with the added REFLEX power spectrum. The continuous line represents
the concordance $\Lambda$CDM structure formation model.}
\label{FIG_PKSUM}
\end{figure}

As mentioned above, hierarchical structure formation predicts that the
amplitudes of the fluctuations should increase in a certain manner
with mass. On scales small compared to the maximum extent of the
survey volume, the fluctuation field roughly follows the superposition
approximation. In this scale range, it is very convinient to test the
mass-dependent amplitude effect with a simple plane wave decomposition
as summarized by the power spectrum $P(k)$\,\footnote{The KL method
would need many modes to test small scales which is presently too
computer-intensiv}. Fig.\,\ref{FIG_DECO} (left) shows that the
observed REFLEX power spectrum of the complete sample is well fit by a
low-density $\Lambda$CDM model.  Comparisons with other hierarchical
scenarios are found to be less convincing (Schuecker et al. 2001c). In
contrast to the `standard CDM' model with $\Omega_{\rm m}=1$, in
low-density (open) CDM models, the epoch of equality of matter and
radiation occure rather late and the growth of structure proceeds over
a somewhat smaller range of redshift, until $(1+z)=\Omega_{\rm
m}^{-1}$. Consequently, the turnover in $P(k)$ is at larger scales,
leaving less power on small scales. The nonzero cosmological constant
of a (flat) $\Lambda$CDM scenario stretches out the time scales of the
model until $(1+z)=\Omega_{\rm m}^{-1/3}$. The differences in the
dynamics of structure growth are thus not very large compared to an
open CDM model and become only important at late stages. Note,
however, that when all models are normalized to the local Universe,
the opposite conclusion is true. The behaviour of the cluster
fluctuation amplitude with mass (X-ray luminosity) for a low-density
CDM model is shown in Fig.\,\ref{FIG_DECO} (right). The predictions
are shown as continuous and dashed lines which nicely follow the
observed trends. The model includes an empirical relation to convert
cluster mass to X-ray luminosity (Reiprich \& B\"ohringer 2002), a
model for quasi-nonlinear and linear structure growth (Peebles 1980),
a biasing model (Mo \& White 1996, Matarrese et al. 1997), and a model
for the transformation of the power spectrum from real space to
redshift space (Kaiser 1987).

However, one could still argue that clusters constitute only a small
population of all cosmological objects visible over a limited redshift
interval, and could therefore not give a representative view of the
goodness of hierarchical structure formation
models. Fig.\,\ref{FIG_PKSUM} summarizes power spectra obtained with
various cosmological tracer objects as compiled by Tegmark \&
Zaldarriaga (2002) including the REFLEX power spectrum. All spectra
are normalized by their respective biasing parameters (if
necessary). The combined power spectrum covers a spatial scale range
of more than four orders of magnitude and redshifts between $z=1100$
(CMB) and $z=0$. The good fit of the $\Lambda$CDM model shows that
this hierarchical structure formation model is really very successful
in describing the clustering properties of cosmological objects. The
following cosmological tests thus assume the validity of this
structure formation model.

\section{Ordinary matter}\label{MATTER}

The observed cosmic density fluctuations are very well summarized by a
low matter density $\Lambda$CDM model (Sect.\,\ref{HIER}). Therefore,
many cosmological tests refer to this structure formation scenario. In
general, baryonic matter, Cold Dark Matter (CDM), primeval thermal
remnants (electromagnetic radiation, neutrinos), and an energy
corresponding to the cosmological constant give the total (normalized)
density of the present Universe, $\Omega_{\rm tot}=\Omega_{\rm
b}+\Omega_{\rm CDM}+\Omega_{\rm r}+\Omega_\Lambda$. The normalized
density of ordinary matter comprises the first three
components. Recent CMB data suggest $\Omega_{\rm tot}=1.02\pm 0.02$
(Spergel et al. 2003), i.e., an effectively flat universe with a
negligible spatial curvature. The same data suggest a baryon density
of $\Omega_{\rm b}h^2=0.024\pm 0.001$ and $h=0.72\pm 0.05$. For our
purposes, the energy density of thermal remnants, $\Omega_{\rm
r}=0.0010\pm0.0005$ (Fukugita \& Peebles 2004), can be neglected,
yielding the present cosmic matter density $\Omega_{\rm m}=\Omega_{\rm
b}+\Omega_{\rm CDM}$. At the end of this section, an estimate of
$\Omega_{\rm r}$ including only the neutrinos is given.

Within this context of the hierarchical structure formation, the
occurence rate of substructure seems to be a useful diagnostic to test
different cosmological parameters because a high merger rate implies a
high $\Omega_{\rm m}$ (e.g., Richstone, Loeb \& Turner 1992, Lacey \&
Cole 1993). However, as mentioned in Sect.\,\ref{GCLST}, the effects
of substructure are difficult to measure and to quantify in terms of
mass so that presently less stringent constraints are attainable (for
a recent discussion see, e.g., Suwa et al. 2003).

A simple though $h$-dependent estimate of $\Omega_{\rm m}$ can be
obtained from the comoving wavenumber of the turnover of the power
spectrum because it corresponds to the horizon length at the epoch of
matter-radiation equality $k_{\rm eq}=0.195\Omega_{\rm m}h^2\,{\rm
Mpc}$ (e.g. Peebles 1993) below which most structure is smoothed-out
by free-streaming CDM particles. A small $\Omega_{\rm m}$ or a small
Hubble constant thus shifts the maximum of $P(k)$ towards larger
scales. The product $\Gamma=\Omega_{\rm m}h$ is referred to as the
shape parameter of the power spectrum. For the REFLEX power spectrum,
the turnover is at $k_{\rm eq}=0.025\pm0.005$ (Fig.\,\ref{FIG_DECO}),
so that for $h=0.72$ a matter density of $\Omega_{\rm m}=0.25\pm0.05$
is obtained. In this case, the shape parameter is $\Gamma=0.18\pm0.03$
which is typical for $\Lambda$CDM.

Cluster abundance measurements are a classical application of galaxy
clusters in cosmology to determine the present density of cosmic
matter, $\Omega_{\rm m}$, either assuming a negligible effect of
$\Omega_\Lambda$ or not. The effective importance of $\Omega_\Lambda$
on geometry and structure growth cannot be neglected for clusters with
$z>0.5$. A related quantity is the variance of the matter fluctuations
in spherical cells with radius $R$ and Fourier transform $W(kR)$:
$\sigma^2(R)=\frac{1}{2\pi^2}\int_0^\infty\,dk\,k^2\,P(k)\,|W(kR)|^2$.
The specific value $\sigma_8$ at $R=8\,h^{-1}\,{\rm Mpc}$
characterizes the normalization of the matter power spectrum
$P(k)$. Recent CMB data suggest $\sigma_8=0.9\pm 0.1$ (Spergel et
al. 2003).

In the following, the abundance of galaxy clusters is used to
determine simultaneously the values of $\Omega_{\rm m}$ and
$\sigma_8$. Early applications of the method can be found in, e.g.,
White, Efstathiou \& Frenk (1993), Eke, Cole \& Frenk (1996), and
Viana \& Liddle (1996) suggesting a stronge degeneracy between
$\Omega_{\rm m}$ and $\sigma_8$ of the form
$\sigma_8=(0.5-0.6)\Omega_{\rm m}^{-0.6}$. To understand this
degeneracy and the high sensitivity of cluster counts on the values of
the cosmological parameters, consider the expected number of clusters
observed at a certain redshift and flux limit,
\begin{equation}\label{ABU1}
dN(z,f_{\rm lim})\,=\,dV(z)\,\int_{M_{\rm lim}(z,f_{\rm lim})}^\infty\,
\,dM\,\,\frac{dn(M,z,\sigma^2(M))}{dM}\,.
\end{equation}
For optically selected samples, the flux limit has to be replaced by a
richness (or optical luminosity) limit. The cosmology-dependency of
$dN$ stems from the comoving volume element $dV$, the mass limit
$M_{\rm lim}$ at a certain redshift, and the shape of the cosmic mass
function $dn/dM$. Three basic cosmological tests are thus applied
simultaneously, which explains the high sensitivity of cluster counts
on cosmology, although sometimes effects related to structure growth
and geometric volume can work against each other (Sect.\,\ref{DE}).

The summation in (\ref{ABU1}) is over cluster mass whereas
observations yield quantities like X-ray luminosity, gas temperature,
richness etc.  The conversion of such observables into mass is the
most crucial step where most of the systematic errors can occure. For
more massive systems, likely contributors to systematic errors are
effects related to cluster merging, substructures, and cooling
cores. Cluster merging increases the gas density and temperature and
thus the X-ray luminosity which increases the detection probablity in
X-rays. The overall statistical effect is difficult to quantify, but
systematic errors in the cosmological parameters on the 20\% level can
be reached (Randall et al. 2002). For less massive systems, further
effects related to additional heat input by AGN, star formation,
galactic winds driven by SNe, etc. lead to deviations from
self-similar expectations (Sect.\,\ref{GCLST}), and increase the
scatter in scaling relations. Such effects are quite difficult to
simulate (e.g., Borgani et al. 2004, Ettori et al. 2004).

Equation\,(\ref{ABU1}) can directly be applied to flux-selected
cluster samples as obtained in X-rays or millimeter wavelengths. The
latter surveys detect clusters via the Sunyaev-Zel'dovich (SZ) effects
(e.g., Birkinshaw, Gull \& Hardebeck 1984, Carlstrom, Holder \& Reese
2002). Here, energy of the ICM electrons is locally transferred
through inverse Compton (Thomson) scattering to the CMB photons so
that the number of photons on the long wavelength side of the Planck
spectrum is depleted. After this blue-shift, each cluster is detected
at wavelengths beyond 1.4\,mm as decrements against the average CMB
background, and at shorther wavelengths as increments. This process
thus measures deviations relative to the actual CMB background and is
thus redshift-independent so that cluster detection does not has to
work against the $(1+z)^4$ Tolman's surface brightness dimming which
is especially important for very distant clusters. Certain blind SZ
surveys are now in preparation (SZ-Array starting 2004; AMI 2004,
APEX-SZ 2005, ACT 2007, SPT 2007 and Planck 2007).  The flux limits in
X-rays and submm allow after some standard corrections a very accurate
determination of the volume accessable by a cluster with certain X-ray
or submm properties.

The detection of clusters in the optical is more complicated (e.g.,
red-sequence method in Gladders, Yee \& Howard 2004, matched filter
method in Postman et al. 1996, Schuecker \& B\"ohringer 1998,
Schuecker, B\"ohringer \& Voges 2004). For the application of
Eq.\,({\ref{ABU1}) to optically selected cluster samples, the mass
limit $M_{\rm min}(z)$ has to be obtained with numerical simulations
in a more model-dependent manner (e.g., Goto et al. 2002, Kim et
al. 2002).

For cosmological tests, the values of the parameters are changed until
observed and predicted numbers of clusters agree. In order to avoid
the evaluation of 3rd and 4th-order statistics in the error
determination, the parameter matrices should be as diagonal as
possible. This can be achieved, when the cluster cell counts are
transformed into the orthonormal base generated by the KL eigenvectors
of the sample correlation matrix (Sect.\,\ref{HIER}). With the REFLEX
sample, the classical $\Omega_{\rm m}$-$\sigma_8$ test was performed
with the KL base (Schuecker et al. 2002, 2003a). The observed
Gaussianity of the matter field directly translates into a
multi-variant Gaussian likelihood function, and includes in a natural
manner a weighting of the squared differences between KL-transformed
observed and modeled cluster counts with the variances of the
transformed counts. Not only the mean counts in the cells but also
their variances from cell to cell depend on the cosmological
model. The KL method thus simultaneously tests both mean counts and
their fluctuations which increases the sensitivity of the method even
more. The method was extensively tested with clusters selected from
the Hubble Volume Simulation. Note that for the application of the KL
method to galaxies of the Sloan Digital Sky Survey (SDSS, Szalay et
al. 2003, Pope et al. 2004) only the fluctuations could be used and
were in fact enough to provide constraints on the 10-percent level.  

A typical result of a cosmological test of $\Omega_{\rm m}$ and
$\sigma_8$ with REFLEX clusters is shown in Fig.\,\ref{FIG_LC}.
\begin{figure}[]
\vspace{0.0cm}
\center{\hspace{-6.5cm}\psfig{figure=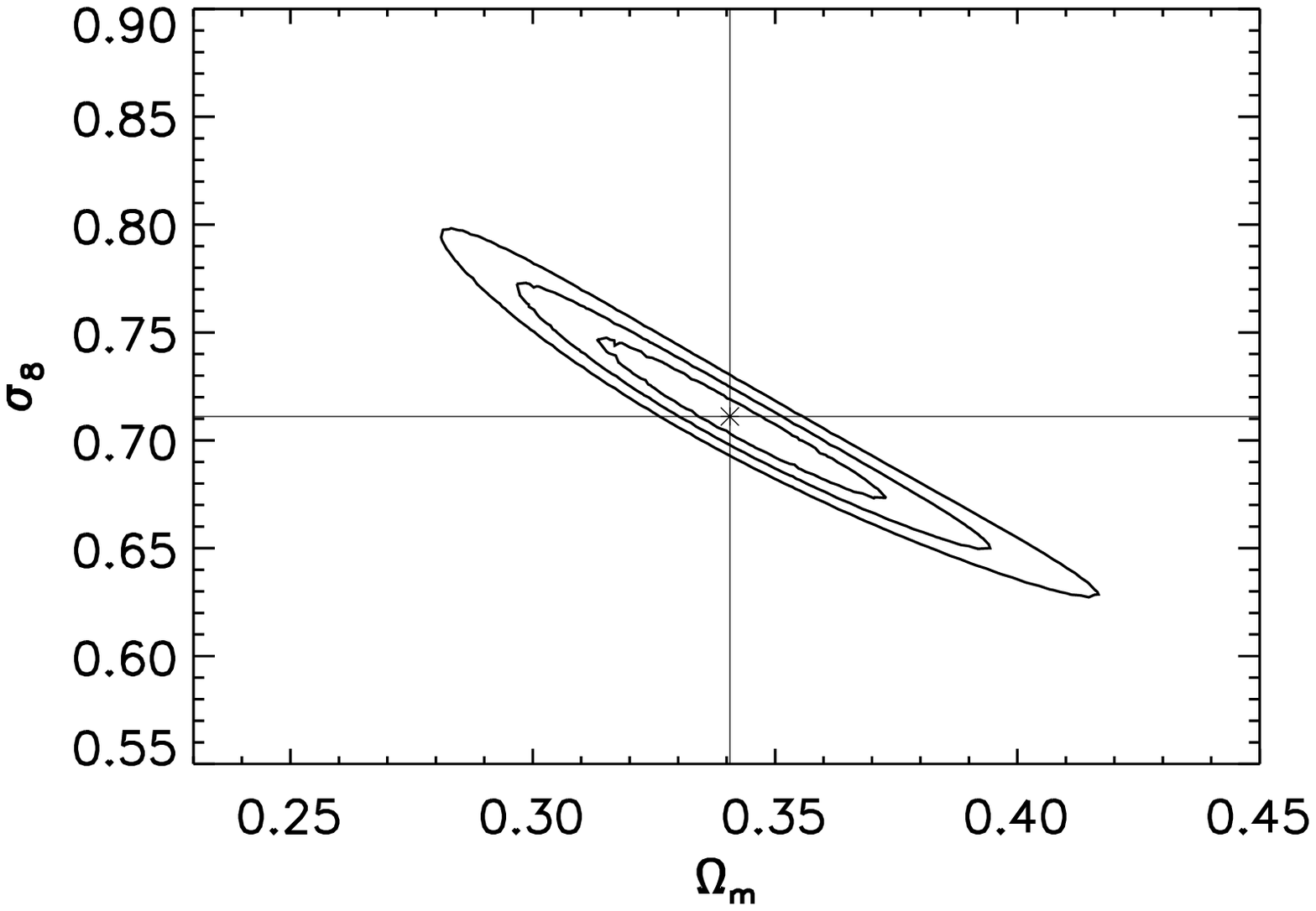,height=4.0cm}}
\vspace{-4.3cm}
\center{\hspace{+4.0cm}\psfig{figure=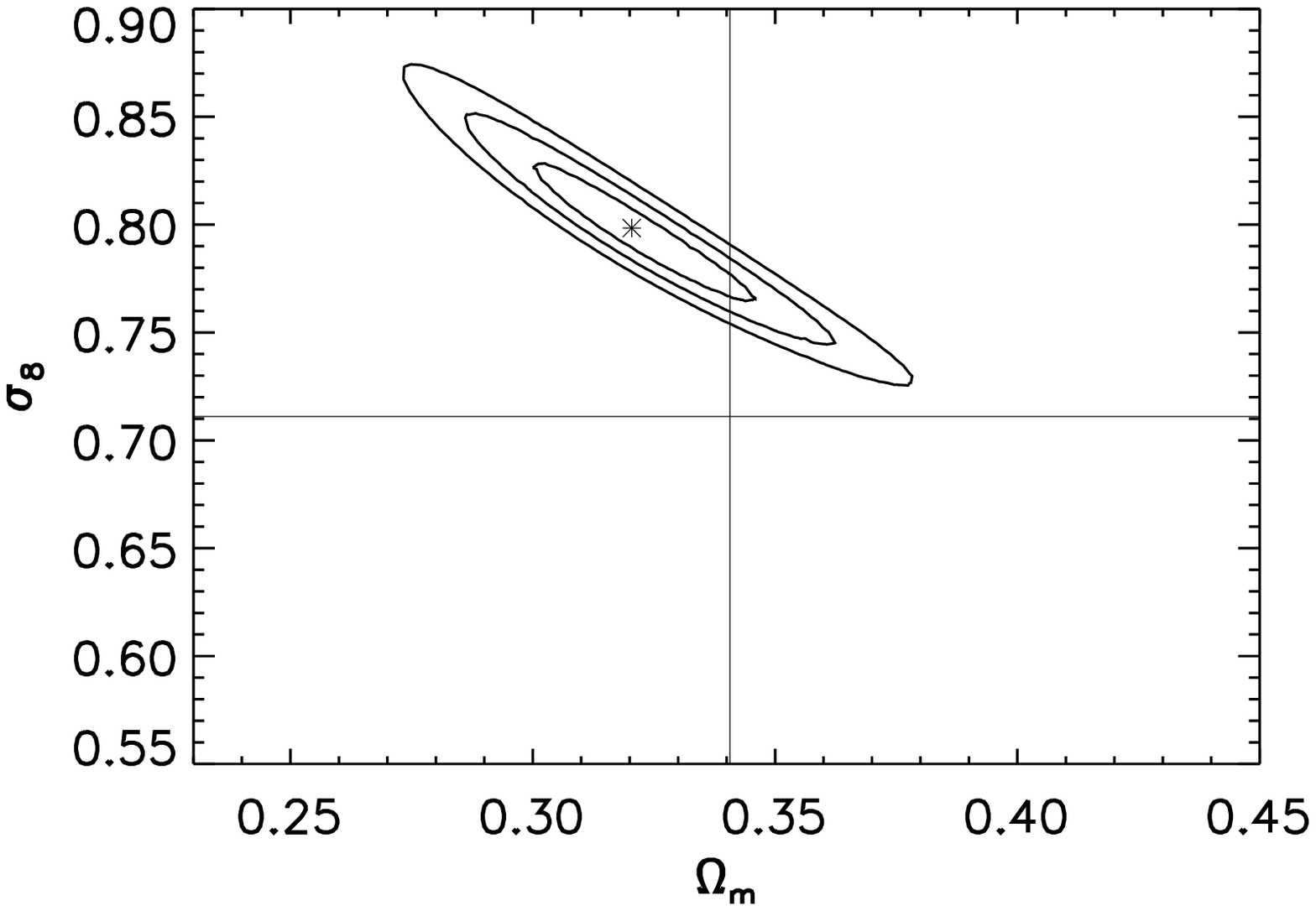,height=4.0cm}}
\vspace{-0.3cm}
\caption[]{\small {\bf Left:} Likelihood contours 
($1-3\sigma$ level for two degrees of freedom) as obtained with the
REFLEX sample. {\bf Right:} Same likelihood contours as left for a
different empirical mass/X-ray luminosity relation.}
\label{FIG_LC}
\end{figure}
Note the small parameter range covered by the likelihood contours and
the residual $\Omega_{\rm m}$-$\sigma_8$ degeneracy: For (flat)
$\Lambda$CDM and low $z$, structure growth is negligible, and the
$\Omega_{\rm m}$-$\sigma_8$ degeneracy is related to the fact that a
small $\sigma_8$ (corresponding to a low-amplitude power spectrum)
yields a small comoving cluster number density, whereas a large
$\Omega_{\rm m}$ (corresponding to a low mass limit $M_{\rm min}$)
yields a large comoving number density. For (flat) $\Lambda$CDM and
high $z$, structure growth and comoving volume do again not strongly
depend on $\Omega_{\rm m}$, but the number of high-$z$ clusters
increases with decreasing $\Omega_{\rm m}$ because for a fixed cluster
number density at $z=0$ the normalization $\sigma_8$ has to be
increased when $\Omega_{\rm m}$ is decreased as shown above. However,
the sensitivity on structure growth becomes apparent once open and
flat models are compared (Bahcall \& Fan 1998).

For the test, further cosmological parameters like the Hubble
constant, the primordial slope of the power spectrum, the baryon
density, the biasing model, and the empirical mass/X-ray luminosity
relation had fixed prior values. The final REFLEX result is obtained
by marginalizing over these parameters and yields the $1\sigma$
corridors $0.28\le\Omega_m\le0.37$ and $0.56\le\sigma_8\le0.80$.

As mentioned above, the largest uncertainty in these estimates comes
from the empirical mass/X-ray luminosity relation obtained for REFLEX
from mainly ROSAT and ASCA pointed observations by Reiprich \&
B\"ohringer (2002) - compare Fig.\,\ref{FIG_LC} left and right. Tests
are in preparation with a four-times larger X-ray cluster sample of
1\,500 clusters combining a deeper version of REFLEX with an extended
version of the cluster catalogue of B\"ohringer et al. (2000) of the
northern hemisphere, plus a more precise M/L-relation obtained over a
larger mass range with the XMM-Newton satellite. Errors below the
10-percent level are expected.

Variants of the cluster abundance method use the X-ray luminosity or
the gas temperature function. For the transition from observables to
mass, often the relations mass-temperature and luminosity-temperature
are used. As an example, Borgani et al. (2001) obtained comparatively
strong constraints using a sample of clusters up to $z=1.27$ yielding
the $1\sigma$ corridors $0.25\le\Omega_{\rm m}\le 0.38$ and $0.61\le
\sigma_8\le 0.72$.

White et al. (1993) pointed out that the matter content in rich nearby
clusters provides a fair sample of the matter content of the
Universe. The ratio of the baryonic to total mass in clusters should
thus give a good estimate of $\Omega_{\rm b}/\Omega_{\rm m}$. The
combination with determinations of $\Omega_{\rm b}$ from BBN
(constrained by the observed abundances of light elements at high $z$)
can thus be used to determine $\Omega_{\rm m}$ (David, Jones, \&
Forman 1995, White \& Fabian 1995, Evrard 1997). Extending the
universality assumption on the gas mass fraction to distant clusters,
Ettori \& Fabian (1999) and later Allen et al. (2002) could show that
at a certain distance from the center of quite relaxed distant
clusters, the observed X-ray gas mass fraction tends to converge to a
universal value. To illustrate the potential power of the method note
that after further corrections, the results obtained by Allen et
al. with only seven apparently relaxed clusters up to $z=0.5$ were
already sensitive enough to constrain the cosmic matter density,
$\Omega_{\rm m}=0.30^{+0.04}_{-0.03}$. Later work includes more
clusters up to $z=0.9$ and cluster abundances from the REFLEX-sample
(B\"ohringer et al. 2004) and the BCS sample (Ebeling et al. 1998),
and yields the $1\sigma$ error corridors $0.25\le
\Omega_{\rm m}\le 0.33$ and $0.66\le\sigma_8\le0.74$ (Allen et
al. 2003). However, the method shares some similarity with the type-Ia
SNe method in the sense that the validity of the gas mass fraction as
a cosmic standard candle especially at high $z$ is mainly based on
observational arguments, partially supported by numerical
simulations. The overlap of the error corridors of the
less-degenerated results of Borgani et al. (2001), Schuecker et
al. (2003a), and Allen et al. (2003) yields our final result
\begin{equation}\label{OMEGA}
\Omega_{\rm m}=0.31\pm0.03\,.
\end{equation}

Other measurements show the $\Omega_{\rm m}$-$\sigma_8$ degeneracy
more pronounced over a larger range. When all measurements are
evaluated at $\Omega_{\rm m}=0.3$, the values of $\sigma_8$ appear
quite consistent at a comparatively low normalization of

\begin{equation}\label{SIGMA8}
\sigma_8\,=\,0.76\,\pm\,0.10\,,
\end{equation}
within the total range $0.5<\sigma_8<1.0$ (data compiled in Henry 2004
from Bahcall et al. 2003, Henry 2004, Pierpaoli et al. 2003, Ikebe et
al. 2002, Reiprich \& B\"ohringer 2002, Rosati et al. 2002, including
Allen et al. 2003 and Schuecker et al. 2003a with small
degeneracies)\footnote{Vauclair et al. (2003) could find a consistent
solution between local and high redshift X-ray temperature
distribution functions and the redshift distributions of distant X-ray
cluster surveys using mass-temperature and luminosity-temperature
relations. Their best model has $\Omega_{\rm m}>0.85$ and
$\sigma_8=0.455$, and the shape parameter, $\Gamma=\Omega_{\rm
m}\,h\approx 0.1$, which implies $h<0.12$, in conflict with many
observations.}.

Recent neutrino experiments are based on atmospheric, solar, reactor,
and accelerator neutrinos. All experiments suggest that neutrinos
change flavour as they travel from the source to the detector. These
experiments give strong arguments for neutrino oscillations and thus
nonzero neutrino rest masses $m_\nu$ (e.g. Ashie et al. 2004 and
references given therein). Further information can be obtained from
astronomical data on cosmological scales. The basic idea is to measure
the normalization of the matter CDM spectrum with CMB anisotropies on
several hundred Mpc scales. This normalization is transformed with
structure growth functions to $8\,h^{-1}\,{\rm Mpc}$ at $z=0$ assuming
various neutrino contributions. This normalization should match the
$\sigma_8$ normalization from cluster counts (e.g., Fukugita, Liu \&
Sugiyama 2000). Recent estimates are obtained by combining CMB-WMAP
data with the 2dFGRS galaxy power spectrum, X-ray cluster gas mass
fractions, and X-ray cluster luminosity functions (Allen, Schmidt \&
Bridle 2003). For a flat universe and three degenerate neutrino
species, they measured the contribution of neutrinos to the energy
density of the Universe, and a species-summed neutrino mass, and their
respective $1\sigma$ errors,
\begin{equation}\label{NEUTR}
\Omega_\nu=0.006\pm0.003\,,\quad\sum_im_i=0.6\pm0.3\,{\rm
eV}\,,
\end{equation}
which formally corresponds to $m_\nu\approx 0.2$\,eV per
neutrino. Their combined analysis yields a normalization of
$\sigma_8=0.74^{+0.12}_{-0.07}$, which is consistent with the recent
measurements with galaxy clusters mentioned above. From CMB, 2dFGRS
and Ly-$\alpha$ forest data, Spergel et al. (2003) obtained the
$2\sigma$ constraint $m_\nu<0.23$\,eV per neutrino. In a similar
analysis including also SDSS galaxy clustering, Seljak et al. (2004)
found $m_\nu<0.13$\,eV for the lightest neutrino (at
$2\sigma$). Estimates from neutrino oscillations suggest $m_\nu\approx
0.05$\,eV for at least one of two neutrino species, consistent with
the Fukugita \& Peebles (2004) estimate given above.

\section{Dark energy}\label{DE}

The present state of the cosmological tests is illustrated in
Fig.\,\ref{FIG_ILLUS} (left). The combination of the likelihood
contours obtained with three different observational approaches
(type-Ia SNe: Riess et al. 2004; CMB: Spergel et al. 2003; galaxy
clusters: Schuecker et al. 2003b) shows that the cosmic matter density
is close to $\Omega_{\rm m}=0.3$, and that the normalized cosmological
constant is around $\Omega_\Lambda=0.7$. This sums up to unit total
cosmic energy density and suggests a spatially flat universe. However,
the density of cosmic matter growths with redshift like $(1+z)^3$
whereas the density $\rho_\Lambda$ related to the cosmological
constant $\Lambda$ is independent of $z$. The ratio
$\Omega_\Lambda/\Omega_{\rm m}$ today is close to unity and must thus
be a finely-tuned infinitesimal constant $\Omega_\Lambda/(\Omega_{\rm
m}(1+z_\infty)^3)$ set in the very early Universe (cosmic coincidence
problem).
\begin{figure}[h]
\vspace{-0.0cm}
\center{\hspace{-6.5cm}\psfig{figure=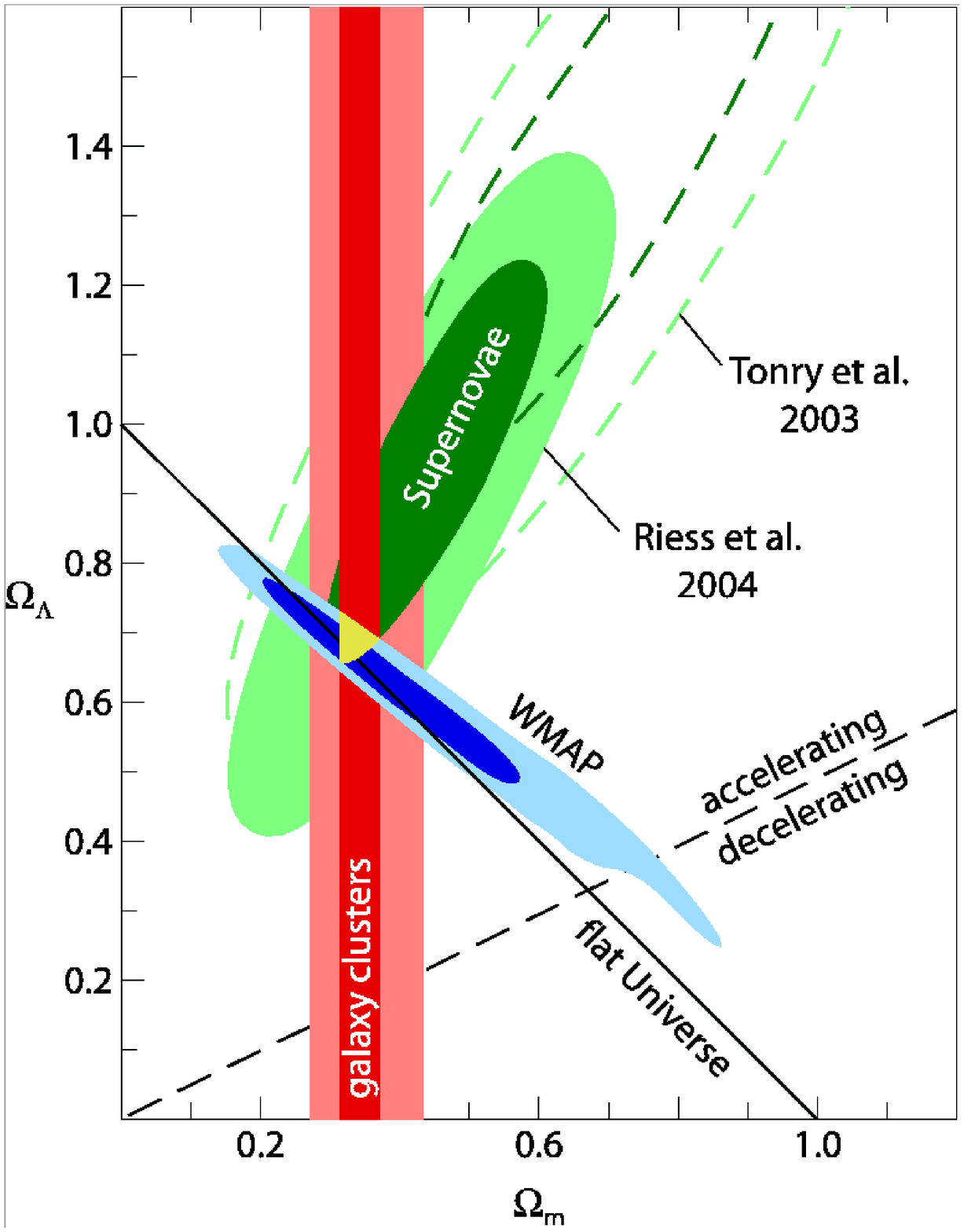,height=6.0cm}}
\vspace{-6.1cm}
\center{\hspace{+5.0cm}\psfig{figure=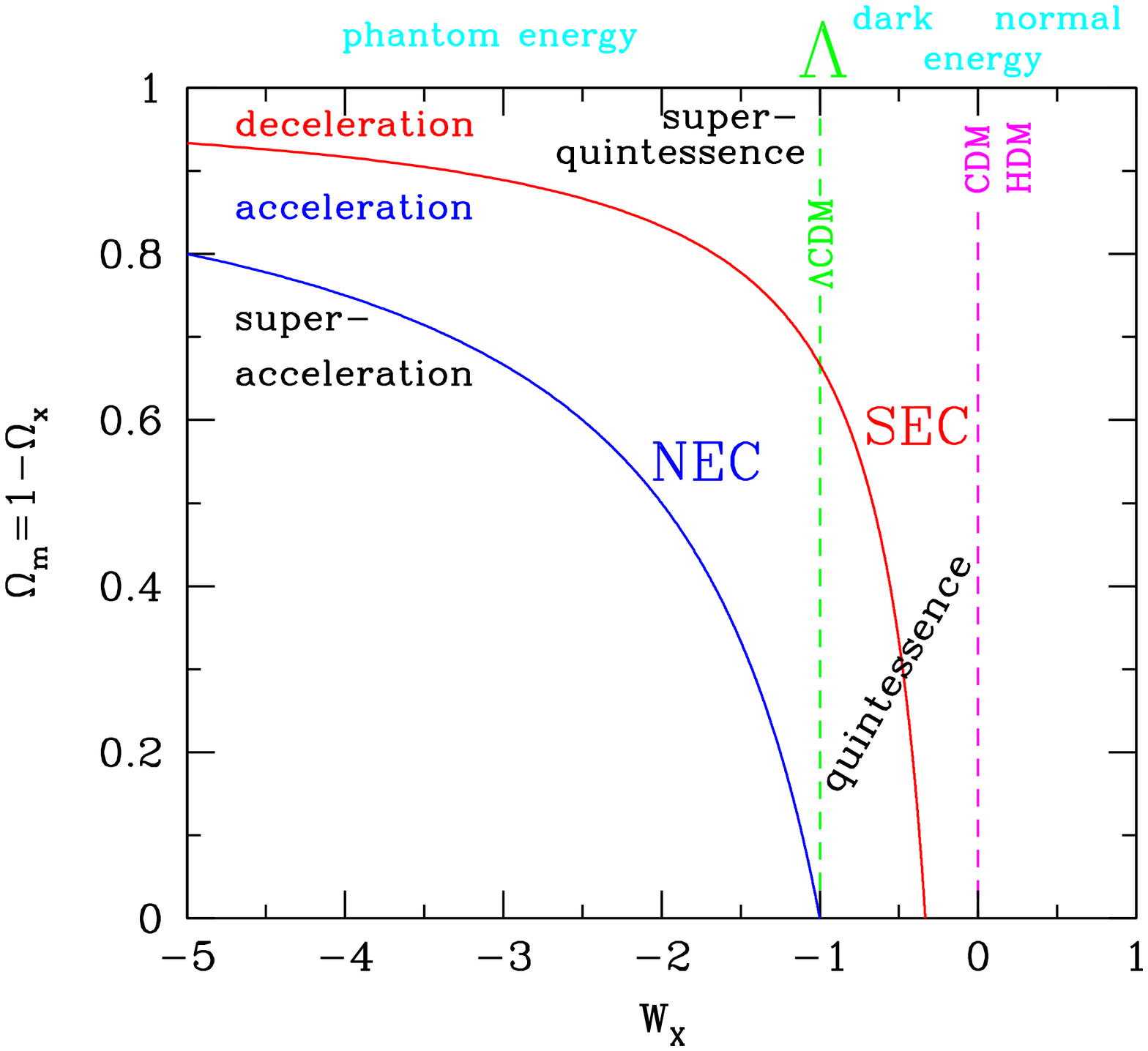,height=6.5cm}}
\vspace{-0.7cm}
\caption[]{\small {\bf Left:} Present situation of 
cosmological tests of the matter density $\Omega_{\rm m}$ and the
normalized cosmological constant $\Omega_\Lambda$ from difference
observational approaches (B\"ohringer, priv. com.). {\bf Right:} Null
Energy Condition (NEC) and Strong Energy Condition (SEC) for a flat FL
spacetime at redshift $z=0$ with negligible contributions from
relativistic particles in the parameter space of the normalized cosmic
matter density $\Omega_{\rm m}$ and the equation of state parameter of
the dark energy $w_{\rm x}$. More details are given in the main text.}
\label{FIG_ILLUS}
\end{figure}
An alternative hypothesis is to consider a time-evolving `dark energy'
(DE), where in Einstein's field equations the time-independent energy
density $\rho_\Lambda$ of the cosmological constant is replaced by a
time-dependent DE density $\rho_{\rm x}(t)$,
\begin{equation}\label{EINSTEIN}
G_{\mu\nu}\,=\,-\frac{8\pi G}{c^4}\,
\left[\,T_{\mu\nu}\,+\,
\rho_{\Lambda\rightarrow{\rm x}}(t)\,c^2\,g_{\mu\nu}\right]\,,
\end{equation}
while assuming that the `true' cosmological constant is either zero or
negligible. Here, $G_{\mu\nu}$ is the Einstein tensor, $T_{\mu\nu}$
the energy-momentum tensor of ordinary matter, and $g_{\mu\nu}$ the
metric tensor. For a time-evolving field (see, e.g., Ratra \& Peebles
1988, Wetterich 1988, Caldwell et al. 1998, Zlatev, Wang \& Steinhardt
1999, Caldwell 2002, recent review in Peebles \& Ratra 2004) the aim
is to understand the coincidence in terms of dynamics. A central
r\^{o}le in these studies is assumed by the phenomenological ratio
\begin{equation}\label{COIN2}
w_{\rm x}=\frac{p_{\rm x}}{\rho_{\rm x} c^2}
\end{equation}
(equation of state) between the pressure $p_{\rm x}$ of the unknown
energy component and its rest energy density $\rho_{\rm x}$. Note that
$w_{\rm x}=-1$ for Einstein's cosmological constant. The resulting
phase space diagram of DE (Fig.\,\ref{FIG_ILLUS}, right) distinguishes
different physical states of the two-component cosmic fluid --
separated by two energy conditions of general relativity (Schuecker et
al. 2003b).

Generally, assumptions on energy conditions form the basis for the
well-known singularity theorems (Hawking \& Ellis 1973), censorship
theorems (e.g. Friedman et al. 1993) and no-hair theorems (e.g. Mayo
\& Bekenstein 1996). Quantized fields violate all local point-wise
energy conditions (Epstein et al. 1965). In the present investigation
we are, however, concerned with observational studies on macroscopic
scales relevant for cosmology where $\rho_{\rm x}$ and $p_{\rm x}$ are
expected to behave classically. Cosmic matter in the form of baryons
and non-baryons, or relativistic particles like photons and neutrinos
satisfy all standard energy conditions. The two energy conditions
discussed below are given in a simplified form (see Wald 1984 and
Barcel\'{o} \& Visser 2001).

The {\it Strong Energy Condition} (SEC): $\rho+3p/c^2\ge0$ {\it and}
$\rho+p/c^2\ge0$, derived from the more general condition
$R_{\mu\nu}v^\mu v^\nu\ge0$, where $R_{\mu\nu}$ is the Ricci tensor
for the geometry and $v^\mu$ a timelike vector. The simplified
condition is valid for diagonalizable energy-momentum tensors which
describe all observed fields with non-zero rest mass and all zero rest
mass fields except some special cases (see Hawking \& Ellis 1973). The
SEC ensures that gravity is always attractive. Certain singularity
theorems (e.g., Hawking \& Penrose 1970) relevant for proving the
existence of an initial singularity in the Universe need an attracting
gravitational force and thus assume SEC. Violations of this condition
as discussed in Visser (1997) allow phenomena like inflationary
processes expected to take place in the very early Universe or a
moderate late-time accelerated cosmic expansion as suggested by the
combination of recent astronomical observations (Fig.\,\ref{FIG_ILLUS}
left). Likewise, phenomena related to $\Lambda>0$ and an effective
version of $\Lambda$ whose energy and spatial distribution evolve with
time ({\it quintessence}: Ratra \& Peebles 1988, Wetterich 1988,
Caldwell et al. 1998 etc.)  are allowed consequences of the breaking
of SEC -- but not a prediction.  However, a failure of SEC seems to
have no severe consequences because the theoretical description of the
relevant physical processes can still be provided in a canonical
manner. Phenomenologically, violation of SEC means $w_{\rm x}<-1/3$
for a {\it single} energy component with density $\rho_{\rm x}>
0$. For $w_{\rm x}\ge -1/3$, SEC is not violated and we have a
decelerated cosmic expansion.

The {\it Null Energy Condition} (NEC): $\rho+p/c^2\ge0$, derived from
the more general condition $G_{\mu\nu}k^\mu k^\nu\ge0$, where
$G_{\mu\nu}$ is the geometry-dependent Einstein tensor and $k^\mu$ a
null vector (energy-momentum tensors as for SEC). Violations of this
condition are recently studied theoretically in the context of
macroscopic traversable wormholes (see averaged NEC: Flanagan \& Wald
1996, Barcel\'{o} \& Visser 2001) and the Holographic Principle
(Sect.\,\ref{CCP}). The breaking of this criterion in a finite local
region would have subtle consequences like the possibility for the
creation of ``time machines'' (e.g. Morris, Thorne
\& Yurtsever 1988). Violating the energy condition in the cosmological
case is not as dangerous (no threat to causality, no need to involve
chronology protection, etc.), since one cannot isolate a chunk of the
energy to power such exotic objects. Nevertheless, violation of NEC on
cosmological scales could excite phenomena like super-acceleration of
the cosmic scale factor (Caldwell 2002). Theoretically, violation of
NEC would have profound consequences not only for cosmology because
all point-wise energy conditions would be broken. It cannot be
achieved with a canonical Lagrangian {\it and} Einstein
gravity. Phenomenologically, violation of NEC means $w_{\rm x}<-1$ for
a {\it single} energy component with $\rho_{\rm x}> 0$. The sort of
energy related to this state of a Friedmann-Robertson-Walker (FRW)
spacetime is dubbed {\it phantom energy} and is described by {\it
super-quintessence} models (Caldwell 2002, see also Chiba, Okabe \&
Yamaguchi 2000). For $w_{\rm x}\ge-1$ NEC is not violated, and is
described by {\it quintessence} or super-quintessence models.

\begin{figure}[]
\vspace{-0.0cm}
\center{\hspace{-1.0cm}\psfig{figure=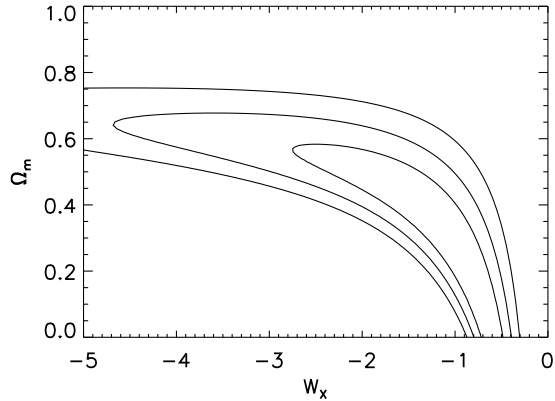,height=6.0cm}}
\vspace{-0.5cm}
\caption[]{\small Likelihood contours ($1-3\sigma$) obtained with the
Riess et al. (1998) sample of type-Ia SNe. The luminosities are
corrected with the $\Delta m_{15}$ method. The equation of state
parameter $w_{\rm x}$ is assumed to be redshift-independent.}
\label{FIG_SN}
\end{figure}

Assuming a spatially flat FRW geometry, $\Omega_{\rm m}+\Omega_{\rm
x}=1$, and $\Omega_{\rm m}\ge 0$ as indicated by the astronomical
observations in Fig.\,\ref{FIG_ILLUS} (left), the formal conditions
for this two-component cosmic fluid translates into $w_{\rm
x}\ge-1/3(1-\Omega_{\rm m})$ for SEC, and $ w_{\rm
x}\ge-1/(1-\Omega_{\rm m})$ for NEC (curved lines in
Fig.\,\ref{FIG_ILLUS} right). These energy conditions, characterizing
the possible phases of a mixture of dark energy and cosmic matter,
thus rely on the precise knowledge of $\Omega_{\rm m}$ and $w_{\rm
x}$. Unfortunately, the effects of $w_{\rm x}$ are not very
large. However, a variety of complementary observational approaches
and their combination helps to reduce the measurement errors
significantly.

The most direct (geometric) effect of $w_{\rm x}$ is to change
cosmological distances. For example, for a spatially flat universe,
comoving distances in dimensionless form,
$a_0r=H_0\int_0^z\frac{dz'}{H(z')}$, are directly related to $w_{\rm
x}$ via
\begin{equation}\label{DIST}
\left[\frac{H(z)}{H_0}\right]^2
 = \Omega_{\rm m}(1+z)^3+(1-\Omega_{\rm m})\exp{\left\{
3\int_0^z[1+w_{\rm x}(z')]\,\,d\ln(1+z')\,
\right\}}\,.
\end{equation}
A less negative $w_{\rm x}$ increases the Hubble parameter and thus
reduces all cosmic distances. In general, $w_{\rm x}$ must evolve in
time. To discuss Eq.\,(\ref{DIST}) in terms of the resulting parameter
degeneracy, let us assume $w_{\rm x}(z)=w_0+w_1\cdot z$ with the
additional constraint that $w_0=-1$ implies $w_1=0$. For this simple
parameterization the same expansion rate at $z$ is obtained when $w_0$
and $w_1$ are related by $w_1=-\frac{\ln(1+z)}{z-\ln(1+z)}(1+w_0)$.
The parameter degeneracy between $w_0$ and $w_1$ is a generic feature
and can be seen in many proposed observational tests. Fortunately, its
slope depends on $z$, so that the degeneracy can be broken with
independent observations covering a large redshift range. Current
observations have not the sensitivity to measure $w_0$ and $w_1$
separately so that basically all published measurements of the
equation of state of the DE are on $w_0$ assuming $w_1=0$. The danger
with this assumption is, however, that if the true $w_1$ would
strongly deviate from zero then the estimated $w_0$ would be biased
correspondingly (Maor et al. 2002). In addition, even when an explicit
redshift dependency of $w_{\rm x}$ could be neglected, some parameter
degeneracy between $\Omega_{\rm m}$ and $w_{\rm x}$ remains as
suggested by Eq.\,(\ref{DIST}) (see Fig.\,\ref{FIG_SN} obtained with
the type-Ia SNe).

\begin{figure}[]
\vspace{0.0cm}
\center{\hspace{-5.5cm}\psfig{figure=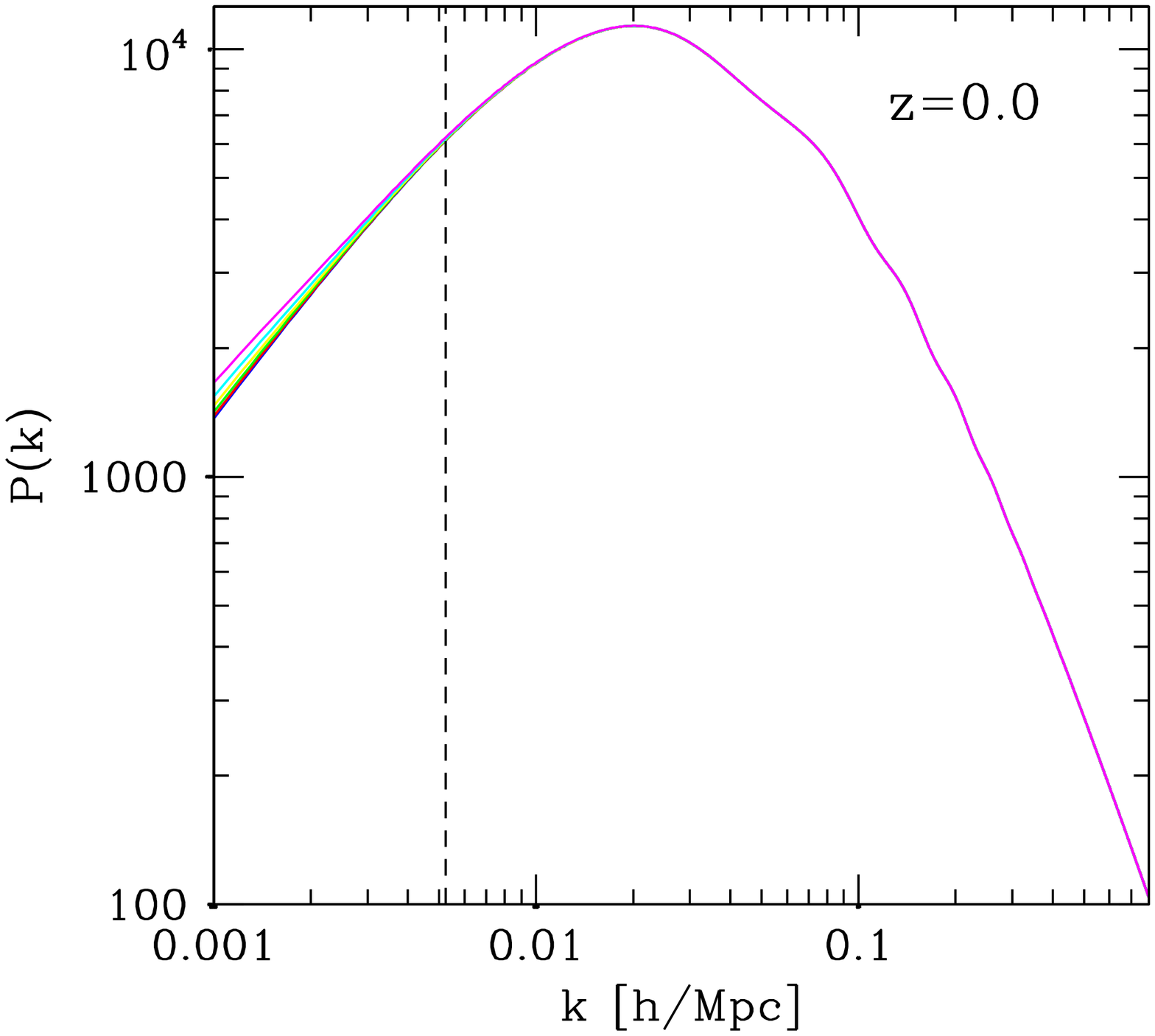,height=5.3cm}}
\vspace{-5.6cm}
\center{\hspace{+6.3cm}\psfig{figure=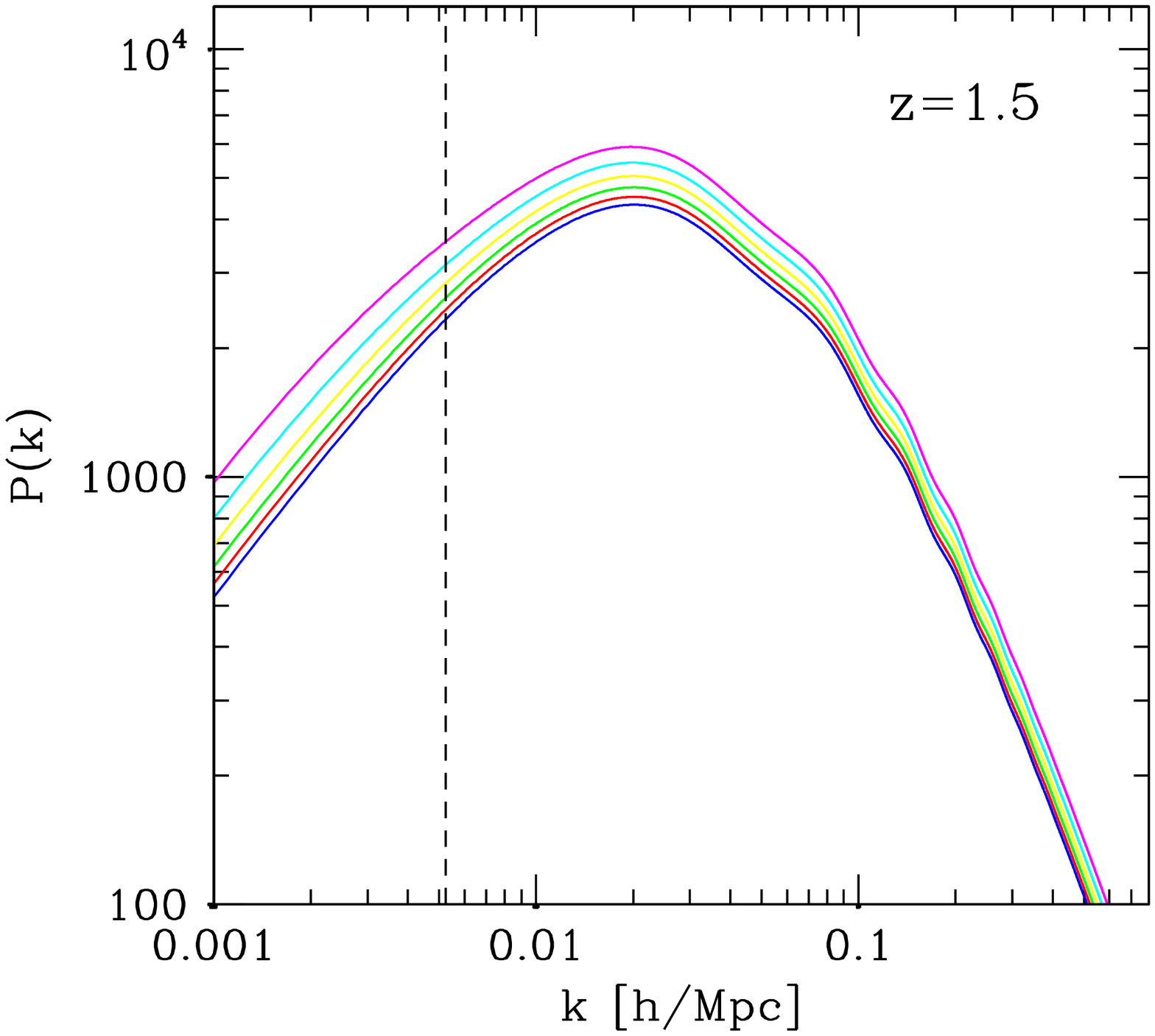,height=5.3cm}}
\vspace{-0.5cm}
\caption[]{\small Evolution of the matter power spectrum for different
redshift-independent equations of state $-1\le w_{\rm x}<0$ of the
DE. The lower curve is for $w_{\rm x}=-1$ and increases in amplitude
with $w_{\rm x}$.}
\label{FIG_PKWZ}
\end{figure}

Structure growth via gravitational instability provides a further
probe of $w_{\rm x}$.  DE, not in form of a cosmological constant or
vacuum energy density, is inhomogenously distributed - a smoothly
distributed, time-varying component is unphysical because it would not
react to local inhomogeneities of the other cosmic fluid and would
thus violate the equivalence principle.  An evolv\-ing scalar field
with $w_{\rm x}<0$ (e.g. quintessence) automatically satisfies these
conditions (Caldwell, Dave \& Steinhardt 1998a). The field is so light
that it behaves relativistically on small scales and
non-relativistically on large scales. The field may develop density
perturbations on Gpc scales where sound speeds $c^2_{\rm s}<0$, but
does not clump on scales smaller than galaxy clusters. Generally,
perturbations come in either linear or nonlinear form depending on
whether the density contrast, $\delta=(\rho/\bar{\rho})-1$, is smaller
or larger than one.

In the linear regime, and when DE is modeled as a dynamical scalar
field, the rate of growth of linear density perturbations in the CDM
is damped by the Hubble parameter, $\delta''_{\rm cdm}+aH\delta'_{\rm
cdm}=4\pi G a^2\delta\rho_{\rm cdm}$ ($a$ means scale factor and prime
derivative with respect to conformal time). This evolution equation
can be solved approximately by $
\frac{d\ln\delta_{\rm cdm}}{d\ln a}\approx 
\left[1+\frac{\rho_{\rm x}(a)}{\rho_{\rm cdm}(a)}\right]^{-0.6}
$ (Caldwell, Dave \& Steinhardt 1998b), provided that $\rho_{\rm
x}<\rho_{\rm r}$ at radiation-matter equality. It is seen that
$\rho_{\rm x}(a)$ and thus a more positive $w_{\rm x}$ delays structure
growth. To reach the same fluctuations in the CDM field, structures
must have formed at higher $z$ compared to the standard CDM model. For
a redshift-independent $w_{\rm x}$, transfer and growth functions can
be found in Ma et al. (1999). The effects of a constant $w_{\rm x}$ on
$P(k)$ are shown in Fig.\,\ref{FIG_PKWZ}. The sensitivity of CMB
anisotropies to $w_{\rm x}$ is limited to the integrated Sachs-Wolfe
effect because $\Omega_{\rm x}$ dominates only at late $z$
(Eq.\,\ref{DIST}). Spergel et al. (2003) showed that the WMAP data
could equally well fit with $\Omega_{\rm m}=0.47$, $h=0.57$, and
$w_{\rm x}=-0.5$ once $w_{\rm x}$ is regarded as a free (constant)
parameter.

In the nonlinear regime, the effects of DE are not very large. For the
cosmological constant, Lahav et al. (1991) used the theory of peak
statistics in Gaussian random fields and linear
gravitational-instability theory in the linear regime and the
spherical infall model to evolve the profiles to the present epoch.
They found that the local dynamics around a cluster at $z=0$ does not
carry much information about $\Lambda$. However, DM haloes have core
densities correlating with their formation epoch. Therefore, when
$w_{\rm x}$ delays structure growth, then DM haloes are formed at
higher $z$ with higher core densities and should thus appear for fixed
mass and redshift more concentrated in $w_{\rm x}> -1$ models compared
to $\Lambda$.  This is reflected in the virial densities of collapsed
objects in units of the critical density shown in Fig.\,\ref{FIG_ADC}
(left). The first semi-analytic computations of a spherical collapse
in a fluid with DE with $-1\le w_{\rm x}<0$ were performed by Wang \&
Steinhardt (1998). Schuecker et al. (2003b) enlarged the range to
$-5<w_{\rm x}<0$, whereas Mota \& van de Bruck (2004) discussed the
spherical collapse for specific potentials of scalar fields. For
recent simulations see Klypin et al. (2003) and Bartelmann et
al. (2004).

\begin{figure}[]
\vspace{0.0cm}
\center{\hspace{-5.8cm}\psfig{figure=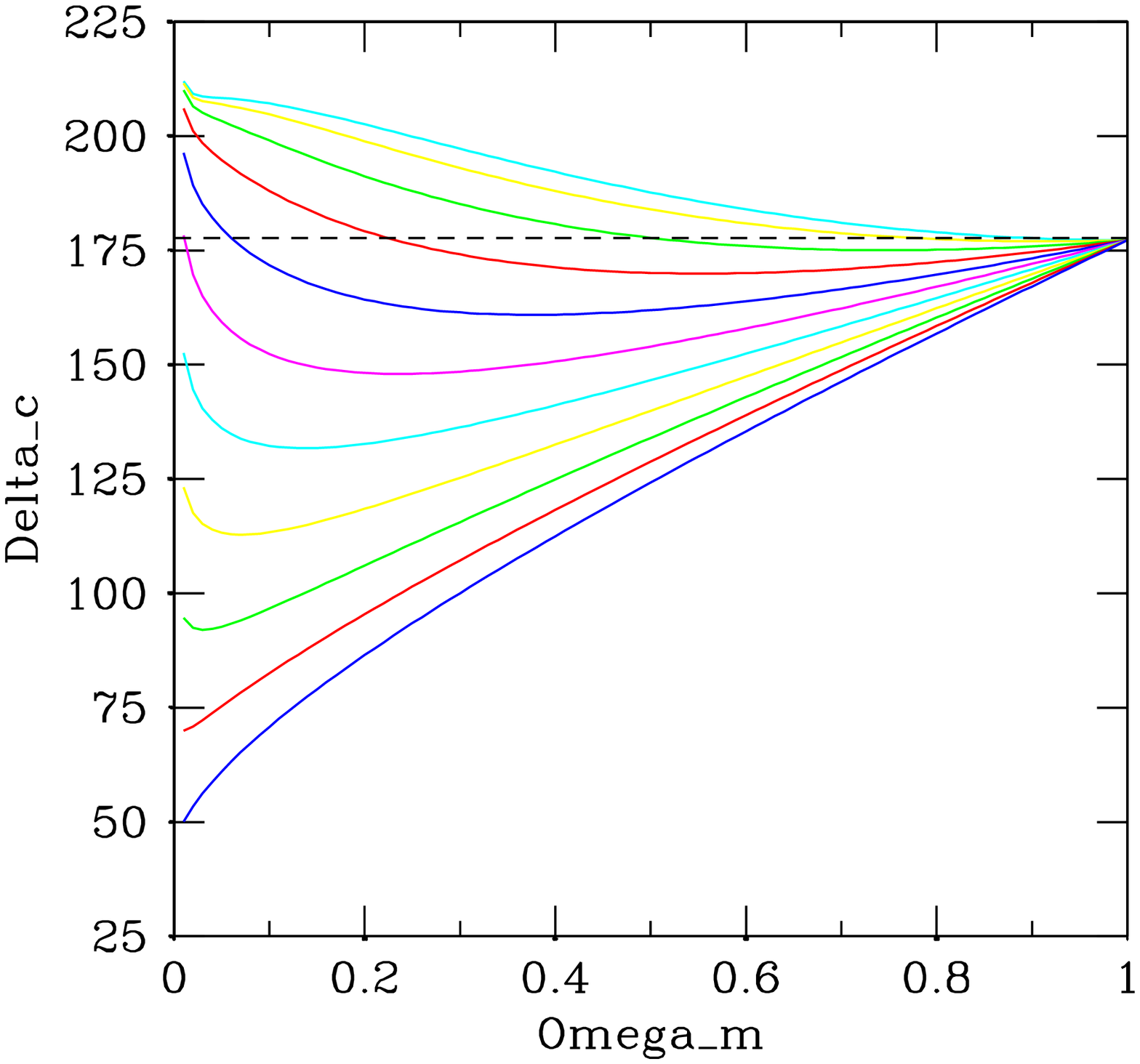,height=6.5cm}}
\vspace{-5.2cm}
\center{\hspace{+5.3cm}\psfig{figure=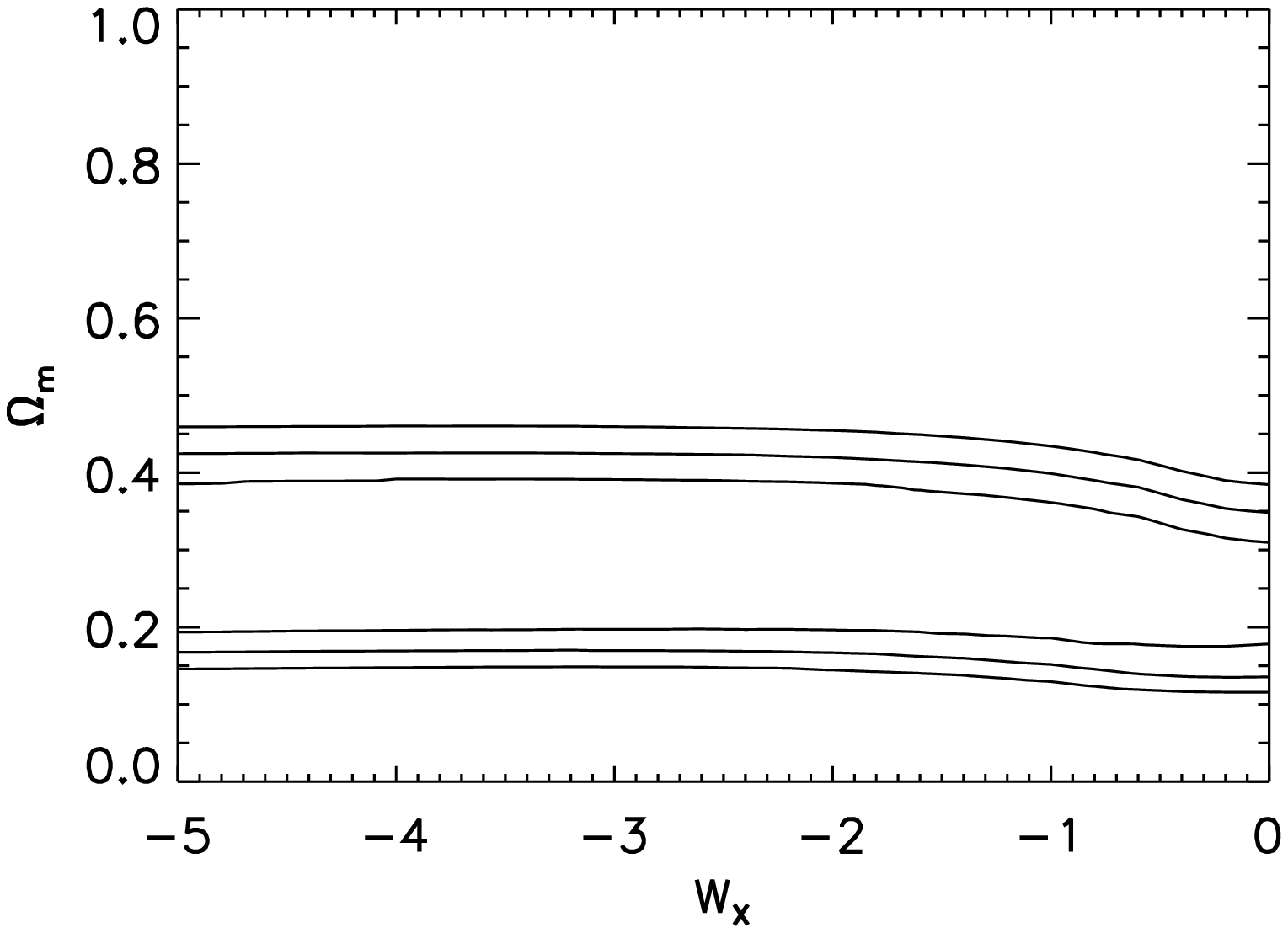,height=4.4cm}}
\vspace{-0.2cm}
\caption[]{\small {\bf Left:} Virial density in units of the critical
matter density for a flat universe as a function of $\Omega_{\rm m}$
and $w_{\rm x}$. The $w_{\rm x}$ values range from $-1$ (lower curve)
to zero (upper curve). {\bf Right:} Likelihood contours
($1$-$3\sigma$) obtained from nearby cluster counts (REFLEX: Schuecker
et al. 2003b) assuming a constant $w_{\rm x}$ and marginalized over
$0.5<\sigma_8<1$. }
\label{FIG_ADC}
\end{figure}

These arguments have to be combined with the general discussion of
Eq.\,(\ref{ABU1}) to understand the sensitivity of cluster counts on
$w_{\rm x}$. Keeping the present-day cluster abundance and lower mass
limit $M_{\rm min}$ in Eq.\,(\ref{ABU1}) fixed, the dominant effect of
$w_{\rm x}$ comes from structure growth and volume (Haiman, Mohr \&
Holder 2001). For a larger $w_{\rm x}$, the DE field delays structure
growth so that the number of distant clusters increases. However, a
large $w_{\rm x}$ yields a small comoving count volume for the
clusters which counteracts the growth effect. The compensation works
mainly at small $z$ and leads to a comparatively small sensitivity of
cluster counts at $z<0.5$ on $w_{\rm x}$.  For $z>0.5$, the effect of
a delayed structure growth starts to dominate and the number of
high-$z$ clusters increases with $w_{\rm x}$. However, the realistic
case is when a redshift and cosmology-dependent lower mass limit is
included. In this case, it could be shown that at high $z$, the
$w_{\rm x}$-dependence of the redshift distribution is mainly caused
by the $w_{\rm x}$-dependence of the lower mass limit in the sense
that a larger $w_{\rm x}$ decreases distances and therefore increases
the number of high-$z$ clusters, whereas at small redshifts no strong
dependency beyond the standard $\Omega_{\rm m}$-$\sigma_8$ degeneracy
remains. The inclusion of a $z$-dependent mass limit does only
slightly damp the sensitivity on $\Omega_{\rm m}$.

This high-$z$ behaviour of the number of clusters is very important
for future planned cluster surveys (e.g. DUO Griffiths et al. 2004)
where in the wide (northern) survey about 8\,000 clusters will be
detected over 10\,000 square degrees on top of the SDSS cap up to
$z=1$, and where in the deep (southern) survey about 1\,800 clusters
will be detected over 176 square degrees up to $z=2$ (if they exists
at such high redshifts). REFLEX has most clusters below $z=0.3$. For a
constant $w_{\rm x}$ the likelihood contours are shown in
Fig.\,\ref{FIG_ADC} (right) as a function of $\Omega_{\rm m}$
(Schuecker et al. 2003b). The effects of yet unknown possible
systematic errors are included by using a very large range of
$\sigma_8$ priors ($0.5<\sigma_8<1.0$). As expected, the $w_{\rm x}$
dependence is very weak.

The past examples (Fig.\,\ref{FIG_SN} and Fig.\,\ref{FIG_ADC} right)
have shown that presently neither SNe nor galaxy clusters alone give
an accurate estimate of the redshift-independent part of $w_{\rm
x}$. This is also true for CMB anisotropies. However, the resulting
likelihood contours of SNe and galaxy clusters appear almost
orthogonal to each other in the high-$w_{\rm x}$ range. Their
combination thus gives a quite strong constraint on both $w_{\rm x}$
and $\Omega_{\rm m}$ (Fig.\,\ref{FIG_SC} left). This is a typical
example of cosmic complementarity which stems from the fact that SNe
probe the homogeneous Universe whereas galaxy clusters test the
inhomogeneous Universe as well. The final result of the combination of
different SNe samples and REFLEX clusters yields the $1\sigma$
constraints $w_{\rm X}=-0.95\pm 0.32$ and $\Omega_{\rm m}=0.29\pm
0.10$ (Schuecker et al. 2003b). Averaging all results obtained with
REFLEX and various SN-samples yields $w_{\rm x}=-1.00^{+0.18}_{-0.25}$
(Fig.\,\ref{FIG_SC} left). The figure shows that the measurements
suggest a cosmic fluid that violates SEC and fulfills NEC. In fact,
the measurements are quite consistent with the cosmological constant
and leave not much room for any exotic types of DE. The violation of
the SEC gives a further argument that we live in a Universe in a phase
of accelerated cosmic expansion.

Ettori, Tozzi \& Rosati (2003) used the baryonic gas mass fraction of
clusters in the range $0.72\le z\le 1.27$ and obtained $w_{\rm x}\le
-0.49$. The combination with SN data yields $w<-0.89$, erroneously
referring to the constraint $w_{\rm x}\ge -1$.  Henry (2004) used the
X-ray temperature function and found $w_{\rm x}=-0.42\pm0.21$,
assuming $w_{\rm x}\ge -1.0$. In a preliminary analysis, Sereno \&
Longo (2004) used angular diameter distance ratios of lensed galaxies
in rich clusters, and shape parameters of surface brightness
distributions and gas temperatures from X-ray data, and obtained
$w_{\rm x}=-0.83\pm0.14$, assuming $w_{\rm x}\ge -1.0$. Rapetti, Allen
\& Weller (2004) combined cluster X-ray gas mass fractions with WMAP
data and obtained the constraints $w_{\rm X}=-1.05\pm 0.11$.  A formal
average of the most accurate und unconstrained $w_{\rm x}$
measurements using galaxy clusters (Schuecker et al. 2003b, Rapetti et
al. 2004) gives
\begin{equation}\label{WVAL}
w_{\rm x}\,=\,-1.00\pm 0.05\,.
\end{equation}

Lima, Cunha \& Alcaniz (2003) give a summary of the results of the
$w_{\rm x}$-$\Omega_{\rm m}$ tests obtained with various methods, all
assuming a redshift-independent $w_{\rm x}$. A clear trend is seen
that $w_{\rm x}>-0.5$ is ruled out by basically all observations. The
large degeneracy seen in Fig.\,\ref{FIG_SN} (left) towards $w_{\rm
x}<-1$ translates into a less well-defined lower bound. Hannestad \&
M\"ortsell (2002) found $w_{\rm x}>-2.7$ by the combination of CMB,
SNe and large-scale structure data. 

Melchiorri et al. (2003) combined seven CMB experiments including WMAP
with the Hubble parameter measurements from the Hubble Space Telescope
and luminosity measurements of type-Ia SNe, and found the 95\%
confidence range $-1.45<w_{\rm x}<-0.74$. If they include also 2dF
data on the large-scale distribution of galaxies they found
$-1.38<w_{\rm x}<-0.82$. More recent measurements support the tendency
that $w_{\rm x}$ is close to the value expected for a cosmological
constant as found by the combination of REFLEX and SN data. Spergel et
al. (2003) used a variety of different combinations between WMAP and
galaxy data and obtained the $1\sigma$ corridor $w_{\rm
X}=-0.98\pm0.12$. Riess et al. (2004) combined data from distant
type-Ia SNe with CMB and large-scale structure data, and found $w_{\rm
x}=-1.02^{+0.13}_{-0.19}$.  Their results are also inconsistent with a
rapid evolution of the DE. Combining Ly-$\alpha$ forest and bias
analysis of the SDSS with previous constraints from SDSS galaxy
clustering, the latest SN and WMAP data, Seljak et al. (2004) obtained
$w_{\rm x}=-0.98^{+0.10}_{-0.12}$ at $z=0.3$ (they also obtained
$\sigma_8=0.90\pm0.03$). A combination of the $w_{\rm x}$ measurements
of REFLEX, Rapetti et al. (2004), Spergel et al. (2003), Riess et
al. (2004), and Seljak et al. (2004) yields $w_{\rm
x}=-0.998\pm0.038$. Independent from this more or less subjective
summary, it is still save to conclude that all recent measurements are
consistent with a cosmological constant, and that the most precise
estimates suggest that $w_{\rm x}$ is very close to $-1$. This points
towards a model where DE behaves very similar to a cosmological
constant, i.e., that the time-dependency of the DE cannot be very
large. In fact, Seljak et al. have also tested $w_{\rm x}$ at $z=1$,
and found $w_{\rm x}(z=1)=-1.03^{+0.21}_{-0.28}$ and thus no
significant change with $z$.

\begin{figure}[h]
\vspace{0.0cm}
\center{\hspace{-5.8cm}\psfig{figure=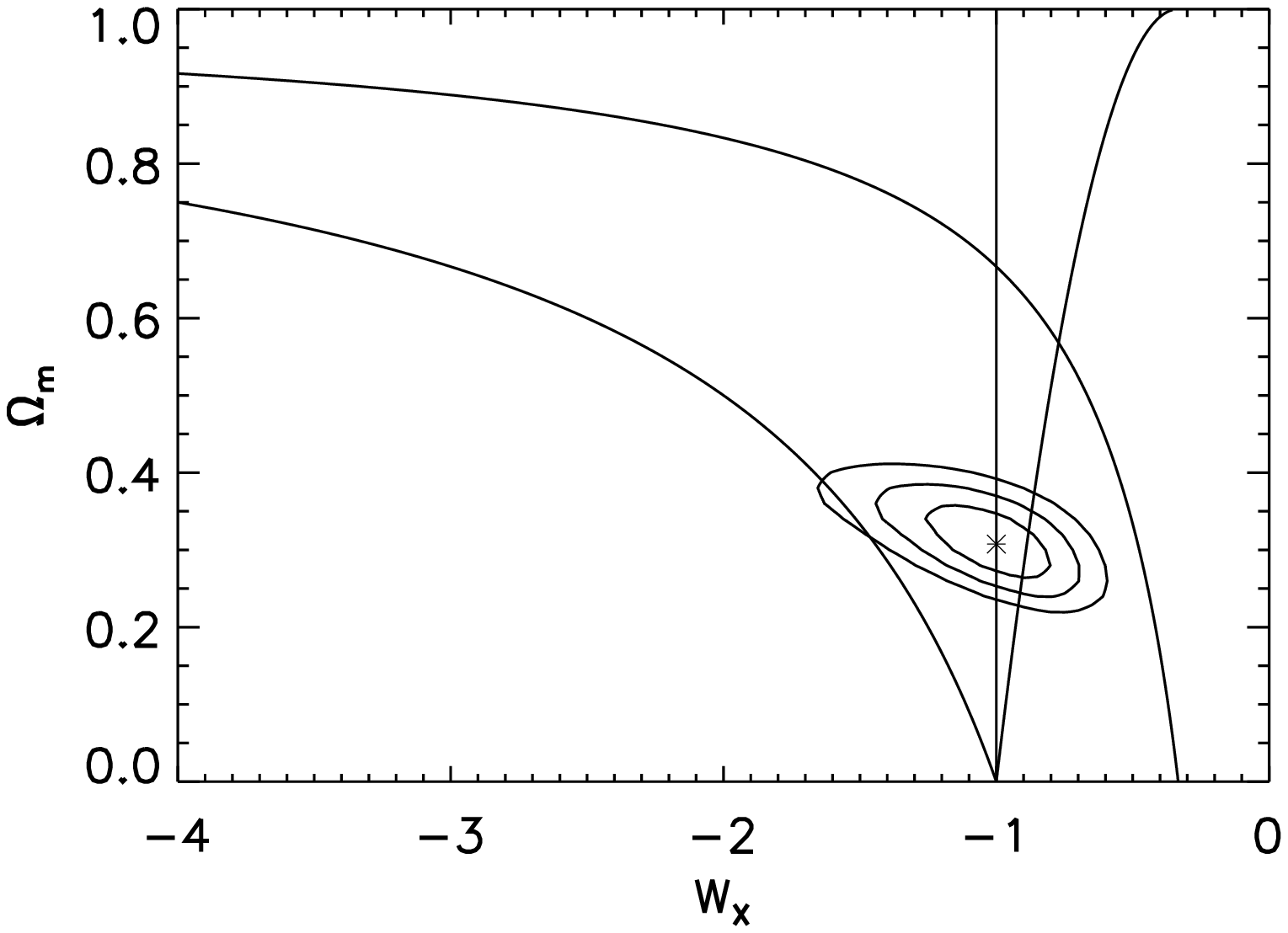,height=5.0cm}}
\vspace{-5.85cm}
\center{\hspace{+5.9cm}\psfig{figure=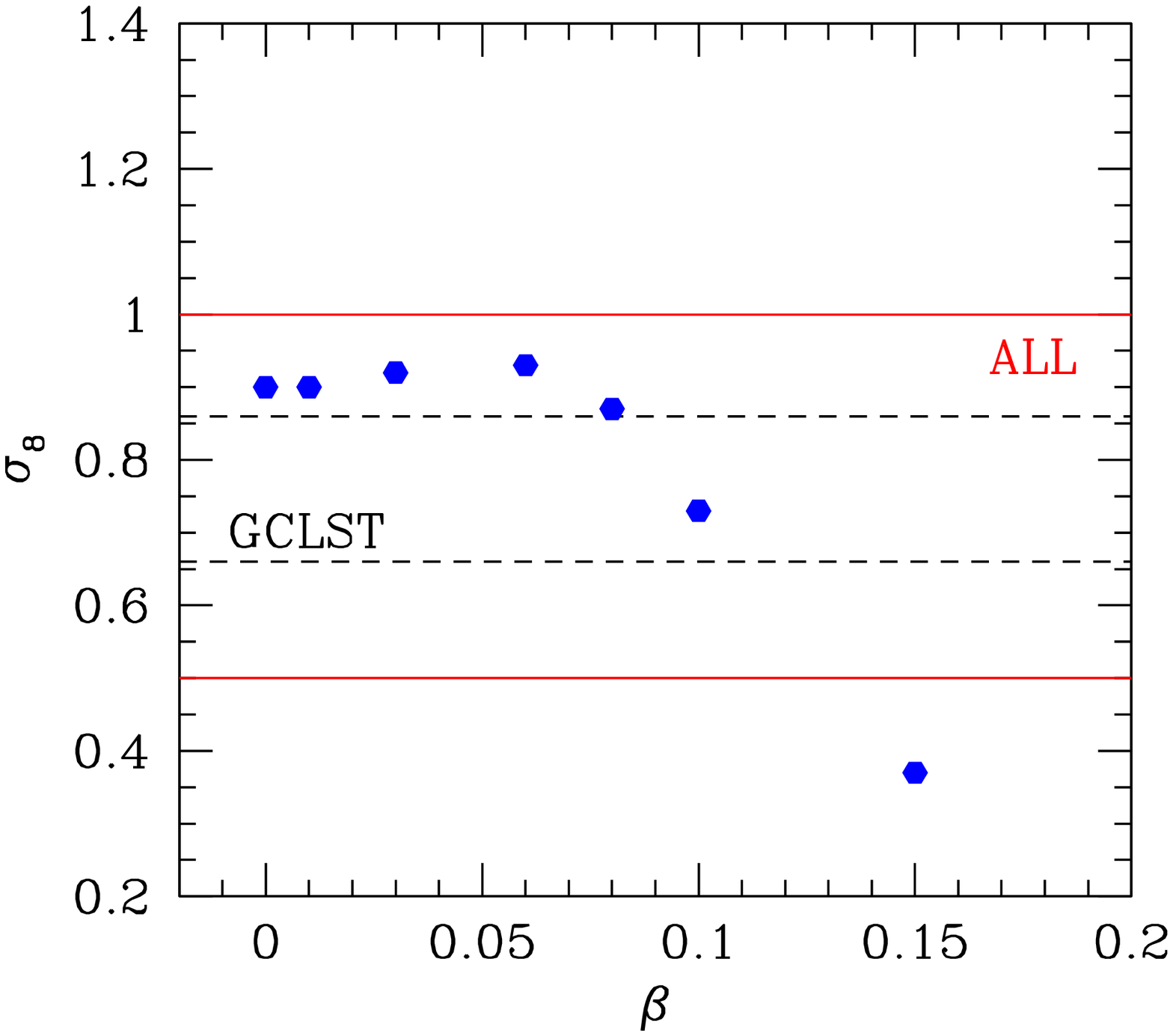,height=5.5cm}}
\vspace{-0.2cm}
\caption[]{\small {\bf Left:} Combination of $w_{\rm x}$ 
measurements based various SN samples and the REFLEX sample assuming a
redshift-independent $w_{\rm x}$. The likelihood contours
($1-3\sigma$) are centred around $w_{\rm x}=-1$ which corresponds to
the cosmological constant (vertical line). The two curved lines
correspond to the SEC (upper line) and the NEC (lower line). The
curved line in the right part of the diagram corresponds to a specific
holographic DE model of Li (2004). {\bf Right:} Normalization
parameter of the matter power spectrum $\sigma_8$ compared to the
coupling strength $\beta$ where $\beta=0$ means no coupling between DE
and DM. The inner region marked by the dashed horizontal lines (GCLST)
marks observational constraints from the scatter of all $\sigma_8$
estimates obtained from galaxy clusters during the past 2 years. The
broader range marked by the continuous horizontal lines (ALL) is a
plausible interval which takes into account also $\sigma_8$
measurements from other observations.}
\label{FIG_SC}
\end{figure}

Cluster abundance measurements have not yet reached the depth to be
very sensitive to the normalized cosmological constant
$\Omega_\Lambda$ or $\Omega_{\rm x}$. The most reliable estimates
todate come from the X-ray gas mass fraction. Vikhlinin et al. (2003)
used the cluster baryon mass as a proxy for the total mass, thereby
avoiding the large uncertainties on the M/T or M/L relations, yielding
with 17 clusters with $z\approx 0.5$ the degeneracy relation
$\Omega_{\rm m}+0.23\Omega_\Lambda=0.41\pm0.10$. For $\Omega_{\rm
m}=0.3$, this would give $\Omega_\Lambda=0.48\pm0.12$. Allen et
al. (2002) obtained with the X-ray gas mass fraction in combination
with the other measurements described above the constraint
$\Omega_\Lambda=0.95^{+0.48}_{-0.72}$. Ettori et al. (2003) obtained
$\Omega_\Lambda=0.94\pm0.30$, and Rapetti et al. (2004)
$\Omega_\Lambda=0.70\pm0.03$. Combining lensing and X-ray data, Sereno
\& Longo (2004) obtained $\Omega_\Lambda=1.1\pm0.2$. The formal
average and $1\sigma$ standard deviation of these measurements is
\begin{equation}\label{Lambda}
\Omega_\Lambda\,=\,0.83\pm0.24\,.
\end{equation}

The last effect of DE and thus $w_{\rm x}$ discussed here is
interesting by its own, but also offers a possibility for cross-checks
of $w_{\rm x}$ measurements. The effect is related to a possible
non-gravitational interaction between DE and ordinary matter
(e.g. Amendola 2000). We showed above (e.g., Eq.\,\ref{WVAL}) that the
most obvious candidate for DE is presently the cosmological constant
with all its catastrophic problems (Sect.\,\ref{CCP}). However, a very
small redshift-dependency of the DE density cannot be ruled out. Based
on this possible residual effect, a further explanation would be a
light scalar (quintessential) field $\phi$ where its potential can
drive the observed accelerated expansion similar as in the de-Sitter
phase of inflationary scenarios. In general, $\phi$ interacts beyond
gravity to baryons and DM with a strength similar to gravity. However,
some (unkown) symmetry could signficantly reduce the interaction
(Carroll 1998) -- otherwise it would have already been detected -- so
that some coupling could remain. The following discussion is
restricted to a possible interaction between DE and DM.

The general covariance of the energy momentum tensor requires the sum
of DM ($m$) and DE ($\phi$) to be locally conserved so that we can
allow for a coupling of the two fluids, e.g., in the simple linear
form,
\begin{eqnarray}\label{COUPL}
T^\mu_{\nu(\phi);\mu}\,&=&\,C(\beta)T_{(m)}\phi_{;\nu}\nonumber\,,\\
T^\mu_{\nu(m);\mu}\,&=&\,-C(\beta)T_{(m)}\phi_{;\nu}\,,
\end{eqnarray}
with the dimensionless coupling constant $\beta$ in
$C(\beta)=\sqrt{\frac{16\pi G}{3c^4}}\beta$, but more complicated
choices are, however, possible. Observational constraints on the
strength of a nonminimal coupling $\beta$ between $\phi$ and DM are
$|\beta|<1$ (Damour et al. 1990). For a given potential $V(\phi)$, the
corresponding equation of motion of $\phi$ can be solved. Amendola
(2000) discussed exponential potentials which yield a present
accelerating phase. A generic result is a saddle-point phase between
$z=10^4$ and $z=1$ where the normalized energy density related to the
scalar field, $\Omega_{\phi}$, is significantly higher compared to
noncoupling models. The saddle-point phase thus leads to a further
suppression of structure growth and thus to smaller $\sigma_8$ (when
the models are normalized with the CMB) compared to noninteracting
quintessence models (Fig.\,\ref{FIG_SC} right). The present
observations appear quite stringent. The X-ray cluster constraint
$\sigma_8=0.76\pm0.10$ (Eq.\,\ref{SIGMA8}) obtained in
Sect.\,\ref{MATTER} suggests a clear detection of a nonminimal
coupling between DE and DM:
\begin{equation}\label{BETA}
\beta\,=\,0.10\,\pm\,0.01\,.
\end{equation}
This would provide an argument that DE cannot be the cosmological
constant because $\Lambda$ cannot couple non-gravitationally to any
type of matter.  In this case, the quite narrow experimental corridor
found for $w_{\rm x}$ (Eq.\,\ref{WVAL}) would be responsible for the
nonminimal coupling. However, a possibly underestimated $\sigma_8$ by
galaxy clusters, and thus no nonminimal couplings and a DE in form of
a cosmological constant seem to provide a more plausible alternative
(see Sect.\,\ref{FUTURE}).

\section{The Cosmological Constant Problem}\label{CCP}

Recent measurements of the equation of state $w_{\rm x}$ of the DE do
not leave much room for any exotic type of DE (Eq.\,\ref{WVAL} in
Sect.\,\ref{DE}). In this section we take the most plausible
assumption that the observed accelerated cosmic expansion is driven by
Einstein's cosmological constant more serious. In this case, we are,
however, confronted with the long-standing cosmological constant
problem (e.g., Weinberg 1989). To some extent also DE models based on
scalar fields suffer on this problem because they have to find a
physical mechanism (symmetry) which makes the value of $\Lambda$
negligible. To illustrate the problem, separate the effectively
observed DE density as usual into a gravitational and
non-gravitational part,
\begin{equation}\label{CCP1}
\rho_{\rm \Lambda}^{\rm eff}=
\rho_{\rm \Lambda}^{\rm GRT}+\rho_{\rm \Lambda}^{\rm VAC}=
10^{-26}\,{\rm kg}\,{\rm m}^{-3}\,,
\end{equation}
for $\Omega_\Lambda=0.7$. The non-gravitational part represents the
physical vacuum. A free scalar field offers a convinient way to get an
estimate of a plausible vacuum energy density. Interpreting this field
as a physical operator and thus constraining it to Heisenberg's
uncertainty relations, quantize the field in the canonical manner. The
quantized field behaves like an infinite number of free harmonic
oscillators. The sum of their zero particle (vacuum) states, up to
the Planck energy, corresponding to a cutoff in physical (not comoving)
wavenumber, is
\begin{equation}\label{CCP2}
\rho_{\rm \Lambda}^{\rm VAC}=\frac{\hbar}{c}\,
\int_0^{E_p/\hbar c}\frac{4\pi k^2 dk}{(2\pi)^3}\frac{1}{2}
\sqrt{k^2+(mc/\hbar)^2}\approx 
10^{+93}\,{\rm kg}\,{\rm m}^{-3}\,,
\end{equation}
for $m=0$. The cosmological constant problem is the extra-ordinary
fine-tuning which is necessary to combine the effectively measured DE
density in Eq.\,(\ref{CCP1}) with the physical vacuum
(\ref{CCP2}). This simple (though quite naive) estimate immediately
shows that something fundamentally has gone wrong with the estimation
of the physical vacuum in Eq.\,(\ref{CCP2}).  An obvious answer is
related to the fact that for the estimation of the physical vacuum,
gravitational effects are completely ignored. One could think of a
quantum gravity with strings. However, present versions of such
theories seem to provide only arguments for a vanishing or a negative
cosmological constant (Witten 2000, but see below).

A hint how inclusion of gravity could effectively work in
Eq.\,(\ref{CCP2}), comes from black hole thermodynamics (Bekenstein
1973, Hawking 1976). Analyzing quantized particle fields in curved but
not quantized spacetimes, it became clear that the information
necessary to fully describe the physics inside a certain region and
characterized by its entropy, increases with the surface of the
region. This is in clear conflict to standard non-gravitational
theories where entropy as an extensive variable always increases with
volume. Non-gravitational theories would thus vastly overcount the
amount of entropy and thus the number of modes and degrees of freedom
when quantum effects of gravity become important. Later studies within
a string theory context could verify a microscopic origin of the black
hole entropy bound (Strominger \& Vafa 1996). Bousso (2002)
generalizes the prescription how entropy has to be determined even on
cosmological scales, leading to the Covariant Entropy Bound. `t Hooft
(1993) and Susskind (1995) elevated the entropy bound as the
Holographic Principle to a new fundamental hypothesis of physics.

A simple intuitive physical mechanism for this holographic reduction
of degrees of freedom is related to the idea that each quantum mode in
Eq.\,(\ref{CCP2}) should carry a certain amount of gravitating
energy. If the modes were packed dense enough, they would immediately
collapse to form a black hole. The reduction of the degrees of freedom
comes from the ignorance of these collapsed states. Later studies of
Cohen, Kaplan \& Nelson (1999), Thomas (2002), and Horvat (2004) made
the exclusion of states inside their Schwarzschild radii more explict
which further strengthen the entropy bound so that a new estimate of
the physical vacuum is
\begin{equation}\label{CCP3}
\rho_{\rm \Lambda}^{\rm HOL}=
\frac{c^2}{8\pi G}\frac{1}{R_{\rm EH}^2}\approx 
3\cdot 10^{-27}\,{\rm kg}\,{\rm m}^{-3}\,,
\end{equation}
where $R_{\rm EH}$ is the present event horizon of the Universe. This
is, however, not a solution of the cosmological constant problem
because gravity and the exclusion of microscopic black hole states were
put in by hand and not in a self-consistent manner by a theory of
quantum gravity. Nevertheless, the similarity of Eqs.\,(\ref{CCP1})
and (\ref{CCP3}) might be taken as a hint that gravitational
hologpraphy could be relevant to find a more complete theory of
physics.

A method to test for consistency of present observations with
gravitational holography, is closely related to the fact that
gravitational holography as tested with the Covariant Entropy Bound on
cosmological scales is based on the validity of the Null Energy
Condition (NEC). However, in contrast to the NEC as discussed in
Sec.\,\ref{DE} for the total cosmic fluid, Kaloper \& Linde (1999)
could show that for the Covariance Entropy Bound each individual
component of the cosmic substratum must obey
\begin{equation}\label{HOLO}
-1\,\le\, w_i\,\le\,+1\,.
\end{equation}
The problematic component is the equation of state of the dark
energy. The observed values summarized in Sect.\,\ref{DE} suggest
$w_{\rm x}=-1.00\pm 0.05$ which is consistent with the bound
(\ref{HOLO}). One can take this as the first consistency test of
probably the most important assumption of the Holographic Principle on
macroscopic scales. However, a direct measurement of {\it cosmological
entropy} on light sheets as defined in Bousso (2002) is still missing.

Li (2004) recently combined holographic ideas with DE to `solve' the
cosmological constant problem. Applying the stronger entropy bound as
suggested by Thomas (1998) and Cohen et al. (1999), and using the
cosmic event horizon as a characteristic scale of the Universe,
accelerating solutions of the cosmic scale factor at low $z$ could be
found together with relations between the density of cosmic matter and
$w_{\rm x}$ as shown in Fig.\,\ref{FIG_SC} (left). This model of
holographic DE appears to be quite consistent with present
observations and was in fact used in Eq.\,(\ref{CCP3}) to estimate the
density of the physical vacuum.

t'Hooft (1993) and Susskind (1995) give arguments suggesting that
M-theory should satisfy the Holographic Principle.  Horava (1999) in
his `conservative' approach to M-theory, defined by specific gauge
symmetries and invariance under spacetime diffeomorphisms and parity,
could show that the entropy bound and thus holography emerges quite
naturally.  Therefore, any astronomical test supporting gravitational
holography more directly or some of its basic assumptions like the NEC
as described above should give important hints towards the development
of a more complete theory of physics.

There is a class of models based on higher dimensions which follow the
Holographic Principle. Brane-worlds emerging from the model of Horava
\& Witten (1996a,b) are phenomenological realizations of M-theory
ideas. Recent theoretical investigations concentrate on the Randall \&
Sundrum (1996a,b) models where gravity is used in an elegant manner to
compactify the extra dimension. Some of these models also follow the
Holographic Principle. Here, matter and radiation of the visible
Universe are located on a $(1+3)$-dimensional brane. Expressed in a
simplified manner, non-gravitational forces, described by open
strings, are attached with their endpoints on branes. Gravity,
described by closed strings, can propagate also into the
$(1+4)$-dimensional bulk and thus `dilutes' differently than Newton or
Einstein gravity. Table-top experiments of classical gravity (and BBN)
confine the size of an extra dimension to $<0.16$\,mm (Hoyle et
al. 2004). Einstein gravity formulated in a five dimensional spacetime
and combined with a five-dimensional cosmic line element carrying the
symmetries of the assumed brane-world, can yield FL-like solutions
with the well-known phenomenology at low $z$ (Binetruy et al. 2000).

\begin{figure}[]
\vspace{-0.0cm}
\center{\hspace{-1.0cm}\psfig{figure=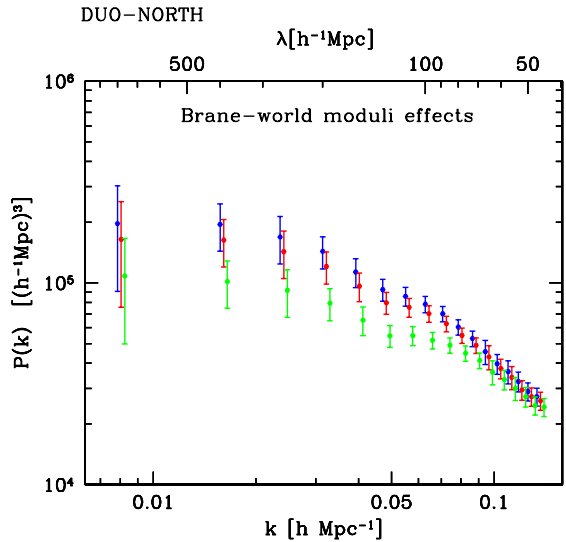,height=8.0cm}}
\vspace{-1.0cm}
\caption[]{\small Predicted cluster power spectra based on matter
power spectra of Rhodes et al. (2003). The effect of the
extra-dimension decreases the $P(k)$ amplitudes at large scales. The
error bars are typical for a DUO-like X-ray cluster survey. In order
to show the differences more clearly, power spectra for each extra
dimension are slidely shifted relative to each other along the
comoving $k$ axis.}
\label{FIG_RHODES}
\end{figure}

The analysis of perturbations in brane-world scenarios is not yet
fully understood (Maartens 2004). Difficulties arise when
perturbations created on the brane propagate into the bulk and react
back onto the brane. Only on large scales are the computations under
control because here the effects of the backreaction are small and can
be neglected. It is thus not yet clear, whether the resulting effects
on the power spectrum described below are mere reflections of such
approximations or generic features of higher dimensions.

Brax et al. (2003) and Rhodes et al. (2003) discussed the effects of
extra dimensions on CMB anisotropies and large scale structure
formation. Models with extra dimensions can at low energies be
described as scalar-tensor theories where the light scalar fields
(moduli fields) couple to ordinary matter in a manner depending on the
details of the higher dimensional theory. An illustration of the
expected effects on the cluster power spectrum is given in
Fig.\,\ref{FIG_RHODES}. The error bars are computed with cluster
samples selected from the Hubble Volume Simulation under the
conditions of the DUO wide survey (P. Schuecker, in prep.). It is seen
that $P(k)$ gets flatter on scales around $300\,h^{-1}\,{\rm Mpc}$
with increasing size of the extra dimension. A careful statistical
analysis shows that more than 30\,000 galaxy clusters are needed to
clearly detect the presence of an extra dimension on scales below
0.16\,mm.

\section{Summary and conclusions}\label{FUTURE}

X-ray galaxy clusters give, in combination with other measurements,
the observational constraints and their $1\sigma$ errors on the matter
density $\Omega_{\rm m}=0.31\pm0.03$, the normalized cosmological
constant $\Omega_\Lambda=0.83\pm0.23$, the normalization of the matter
power spectrum $\sigma_8=0.76\pm0.10$, the neutrino energy density
$\Omega_\nu=0.006\pm0.003$, the equation of state of the DE $w_{\rm
x}=-1.00\pm0.05$, and the linear interaction $\beta=0.10\pm0.01$
between DE and DM. These estimates suggest a spatially flat universe
with $\Omega_{\rm tot}=\Omega_{\rm m}+\Omega_\Lambda=1.14\pm0.24$, as
assumed in many cosmological tests based on galaxy clusters.

They do, however, not provide an overall consistent physical
interpretation. The problem is related to the low $\sigma_8$ which
leads to an overestimate of the neutrino mass compared to laboratory
experiments and to an interaction between DE and DM. Such a high
interaction is not consistent with a DE with $w_{\rm x}=-1.00\pm0.05$
because the latter indicates that DE behaves quite similar to a
cosmological constant which cannot exchange energy beyond gravity. 

A more convincing explanation is that $\sigma_8=0.76$ should be
regarded as a lower limit so that DE would be the cosmological
constant without any nonminimal couplings. Systematic underestimates
of $\sigma_8$ by 10-20\% are not unexpected from recent simulations
(e.g., Randall et al. 2002, Rasia et al. 2004). Present data do not
allow any definite conclusion, especially in the light of the
partially obscured effects of non-gravitational processes in galaxy
clusters and because of our ignorance of a possible time-dependency of
$w_{\rm x}$. However, the inclusion of further parameters obviously
improves our abilities for consistency checks.

Energy conditions form the bases of many phenomena related to gravity
and holography. M-theory should also come holographic, as well as
brane-world gravity as a phenomenological realization of M-theory
ideas. Tests of the resulting cosmologies will in the end confront
alternative theories of quantum gravity. Observational tests on
cosmological scales as illustrated by the effects of an
extra-dimension on the cluster power spectrum probably need the
`ultimate' cluster survey, i.e. a census of possibly all $10^6$ rich
galaxy clusters which might exist down to redshifts of $z=2$ in the
visible Universe.\\ \\

{\it Acknowledgements: I would like to thank Hans B\"ohringer and the
REFLEX team for our joint work on galaxy clusters and cosmology.}

\subsection*{References}

{\small

\bref
Allen, S.W., Schmidt, R.W., Fabian, A.C., 2002, MNRAS, 334, L11

\bref
Allen, S.W., Schmidt, R.W., Fabian, A.C., Ebeling, H., 2003, MNRAS,
344, 43

\bref
Allen, S.W., Schmidt, R.W., Bridle, S.L., 2003, MNRAS, 346, 593

\bref
Allen, S.W., Schmidt, R.W., Ebeling, H., Fabian, A.C., van Speybroeck,
L., 2004, MNRAS, 353, 457

\bref
Amendola, L., 2000, PhRvD, 62, 043511

\bref
Amossov, G., Schuecker, P., 2004, A\&A, 421, 425

\bref
Ashie, Y., et al., 2004, PhRvL, 93, 101801

\bref
Bahcall, N.A., 1999, in Formation of structure in the universe,
eds. A. Dekel \& J.P. Ostriker, Cambridge Univ. Press, Cambridge,
p.\,135

\bref
Bahcall, N.A., Fan, X., 1998, ApJ, 504, 1

\bref
Bahcall, N.A., et al. 2003, ApJ, 585, 182

\bref
Barcel\'{o}, C., \& Visser, M., 2001, PhLB, 466, 127

\bref
Bartelmann, M., Dolag, K., Perrotta, F., Baccigalupi, C., Moscardini,
L., Meneghetti, M., Tormen, G., 2004, astro-ph/0404489

\bref
Bekenstein, J., 1973, PhRvD, 9, 3292

\bref
Binetruy, P., Deffayet, C., Ellwanger, U., Langois, D., 2000, PhLB,
477, 285

\bref
Birkingshaw, M., Gull, S.F., Hardebeck, H., 1984, Natur, 309, 34

\bref
B\"ohringer, H., Voges, W., Fabian, A.C., Edge, A.C., Neumann, D.M.,
1993, MNRAS, 264, L25

\bref
B\"ohringer, H., et al., 2000, ApJS, 129, 35

\bref
B\"ohringer, H., et al. 2001, A\&A, 369, 826

\bref
B\"ohringer, H., et al. 2002, ApJ, 566, 93

\bref
B\"ohringer, H., et al. 2004, A\&A, 425, 367

\bref
Bond, J.R., 1995, PhRvL, 74, 4369

\bref
Borgani, S., Guzzo, L., 2001, Natur, 409, 39

\bref
Borgani, S., Rosati, P., Tozzi, P., Norman, C., 1999, ApJ, 517, 40

\bref
Borgani, S., et al. (2001), ApJ, 561, 13

\bref
Borgani, S., Murane, G., Springel, V., Diaferio, A., Dolag, K.,
Moscardini, L., Tormen, G., Tornatore, L., Tozzi, P., 2004, MNRAS,
348, 1078

\bref
Bousso, R., 2002, RvMP, 74, 825

\bref
Brax, Ph., van de Bruck, C., Davis, A.-C., Rhodes, C.S., 2003, PhRvD,
67, 023512

\bref
Briel, U.G., Henry, J.P., B\"ohringer, H., 1992, A\&A, 259, L31

\bref
Briel, U.G., Finoguenov, A., Henry, J.P., 2004, A\&A, 426, 1

\bref
Burles, S., Nollett, K.M., Turner, M.S., 2001, ApJL, 552, L1

\bref
Caldwell, R.R., 2002, PhLB, 545, 17

\bref
Caldwell, R.R., Dave, R., Steinhardt, P.J., 1998a, PhRvL,
80, 1582

\bref
Caldwell, R.R., Dave, R., Steinhardt, P.J., 1998b, Ap\&SS, 261, 303

\bref
Carlstrom, J.E., Holder, G.P., Reese, E.D., 2002, ARAA, 40, 643

\bref
Carroll, S.M., 1998, PhRvL, 81, 3067

\bref
Chaboyer, B., Krauss, M., 2002, ApJL, 567, L45

\bref
Chiba, T., Okabe, T., Yamaguchi, M., 2000, PhRvD, 62,
023511

\bref
Cohen, A.G., Kaplan, D.B., Nelson, A.E., 1999, PhRvL, 82, 4971

\bref
Collins, C.A., et al. 2000, MNRAS, 319, 939

\bref
Cruz, M., Martinez-Gonzalez, E., Vielva, P., Cayon, L., 2004, MNRAS
(in press)

\bref
Damour, T., Gibbons, G.W., Gundlach, C., 1990, PhRvL, 64, 123

\bref
David, L.P., Jones, C., Forman, W., 1995, ApJ, 445, 578

\bref
Ebeling, H., Edge, A.C., B\"ohringer, H., Allen, S.W., Crawford, C.S.,
Fabian, A.C., Voges, W., Huchra, J.P., 1998, MNRAS, MNRAS, 301, 881

\bref
Edge, A.C., 2004, in Clusters of Galaxies, eds. J.S. Mulchaey,
A. Dressler, and A. Oemler, Cambridge Univ. Press, Cambridge, p.\,58

\bref
Efstathiou, G., Frank, C.S., White, S.D.M., Davis, M., 1988, MNRAS,
235, 715

\bref
Eke, V.R., Cole, S., Frenk, C.S., 1996, MNRAS, 282, 263

\bref
Epstein, H., Glaser, V., Jaffe, A., 1965, NCim, 36, 2296

\bref
Ettori, S., Fabian, A.C., 1999, MNRAS, 305, 834

\bref
Ettori, S., Tozzi, P., Rosati, P., 2004, A\&A, 398, 879

\bref
Ettori, S., et al. 2004, MNRAS, 354, 111

\bref
Evrard, A.E., 1997, MNRAS, 292, 289

\bref
Evrard, A.E., Metzler, C.A., Navarro, J.F., 1996, ApJ, 469, 494

\bref
Fabian, A.C., et al. 2000, MNRAS, 318, L65

\bref
Fabian, A.C., Sander, J.S., Allen, S.W., Crawford, C.S., Iwasawa, K.,
Johnstone, R.M., Schmidt, R.W., Taylor, G.B., 2003, MNRAS, 344, L43

\bref
Feretti, L., Gioia, I.M., Giovannini, G., I. Gioia 2002, Merging
processes in galaxy clusters, Eds., Astrophysics and Space Science
Library, Vol. 272, Kluwer Academic Publisher, Dordrecht

\bref
Flanagan, \'{E}.\'{E}., Wald, R.M., 1996, PhRvD, 54, 6233

\bref
Forman, W., et al., 2003, ApJ (submitted), astro-ph/0312576

\bref
Friedman, J.L., Schleich, K., Witt, D.M., 1993,
PhRvL, 71, 1486

\bref
Fukuda, Y., et al., 1998, PhRvL, 81, 1562

\bref
Fukugita, M., Liu, G.-C., Sugiyama, N., 2000, PhRvL, 84, 1082

\bref
Fukugita, M., Peebles , J.P.E. 2004, astro-ph/0406095

\bref
Gladders, M.D., Yee, D., Howard, K.C., 2004, ApJS (in press),
astro-ph/0411075

\bref
Goto, T., et al. 2002, AJ, 123, 1807

\bref
Griffiths, R.E., Petre, R., Hasinger, G., et al. 2004, in Proc. SPIE
conference (submitted)

\bref
Haiman, Z., Mohr, J.J.,  Holder, G.P., 2001, ApJ, 553, 545

\bref
Hawking, S.W.,  Penrose, R., 1970, in Proc. of the Roycal Society of
London. Series A. Mathematical and Physical Sciences. Vol. 314, Issue
1519, p.\,529

\bref
Hawking, S.W., Ellis, G.F.R., 1973, The large scale
structure of space-time, Cambridge Monographs on Mathematical Physics,
Cambridge Univ. Press, London

\bref
Hawking, S.W., 1976, PhRvD, 13, 191

\bref
Henry, J.P. 2004, ApJ, 609, 603

\bref
Horava, P., 1999, PhRvD, 59, 046004

\bref
Horava, P., Witten, E., 1996a, NuPhB, 460, 506

\bref
Horava, P., Witten, E., 1996b, NuPhB, 475, 94

\bref
Horvat, R., 2004, PhRvD, 70, 087301

\bref
Hoyle, C.D., Kapner, D.J., Heckel, B.R., Adelberger, E.G., Gundlach,
J.H., Schmidt, U., Swanson, H.E., 2004, PhRvD, 70, 042004

\bref
Ikebe, Y., Reiprich, T.H., B\"ohringer, H., Tanaka, Y., Kitayama, T.,
2002, A\&A, 383, 773

\bref
Jenkins, A., Frenk, C.S., White, S.D.M., et al. 2001, MNRAS, 321, 372

\bref
Kaiser, N., 1984, ApJL, 284, L9

\bref
Kaiser, N., 1986, MNRAS, 222, 323

\bref
Kaiser, N., 1987, MNRAS, 227, 1

\bref
Kaiser, N., Squires, G., 1993, ApJ, 404, 441

\bref
Kaloper, N., Linde, A., 1999, PhRvD, 60, 103509

\bref
Komatsu, E., et al., 2003, ApJS, 148, 119

\bref
Kim, R.S.J., et al., 2002, AJ, 123, 20

\bref
Klypin, A., Maccio, A.V., Mainini, R., Bonometto, S.A., 2003, ApJ,
599, 31

\bref
Lacey, C.G., Cole, S.M., 1993, MNRAS, 262, 627

\bref
Lacey, C.G., Cole, S.M., 1994, MNRAS, 271, 676

\bref
Lahav, O., Rees, M.J., Lilje, P.B., Primack, J.R., 1991, MNRAS, 251,
128

\bref
Li, M., 2004, Phys. Lett. B (submitted), astro-ph/0403127

\bref
Lima, J.A.S., Cunha, J.V., Alcaniz, J.S., 2003, PhRvD, 68, 023510

\bref
Ma, C.-P., Caldwell, R.R., Bode, P., Wang, L., 1999, ApJ,
521, L1

\bref
Maartens, R., 2004, LRR, 7, 7

\bref
Maor, I., Brustein, R., McMahon, J., Steinhardt, P.J., 2002, PhRvD,
65, 123003

\bref
Matarrese, S., Coles, P., Lucchin, F., Moscardini, L., 1997, MNRAS,
286, 115

\bref
Mather, J.C., et al. 1990, ApJ, 354, L37

\bref
Mayo, A.E., Bekenstein, J.D., 1996, PhRvD, 54, 5059

\bref
Melchiorri, A., Mersini, L., \"Odman, C.J., Trodden, M., 2003,
PhRvD, 68, 043509

\bref
Mo, H.J., White, S.D.M, 1996, MNRAS, 282, 347

\bref
Morris, M.S., Thorne, K.S., Yurtsever, U., 1988,
PhRvL, 61, 1446

\bref
Mota, D.F., van de Bruck, C., 2004, A\&A, 421, 71

\bref
Peebles, P.J.E., 1980, The Large-Scale Structure of the Universe, 
Princeton Univ. Press, Princeton

\bref
Peebles, P.J.E., 1993, Principles of Physical Cosmology,
Univ. Press, Princeton, Princeton

\bref
Peebles, P.J.E., Ratra, B., 2004 RvMP, 75, 559

\bref
Pierpaoli, E., Borgani, S., Scott, D., White, M., 2003, MNRAS, 242,
163

\bref
Ponman, T.J., Cannon, D.B., Navarro, J.F., 1999, Natur, 397, 135

\bref
Pope, A.C., et al. 2004, ApJ, 607, 655

\bref
Postman, M., Lubin, L.M., Gunn, J.E., Oke, J.B., Hoessel, J.G.,
Schnieder, D.P., Christensen, J.A., 1996, AJ, 111, 615

\bref
Randall, L., Sundrum, R., 1996a, PhRvL, 83, 3370

\bref
Randall, L., Sundrum, R., 1996b, PhRvL, 83, 4690

\bref
Randall, S.W., Sarazin, C.L., Ricker, P.M., 2002, ApJ, 577, 579

\bref
Rapetti, D., Allen, S,W, Weller, J., 2004, MNRAS (submitted),
astro-ph/0409574

\bref
Rasia, E., Mazzotta, P., Borgani, S., Moscardini, L., Dolag, K.,
Tormen, G., Diaferio, A., Murante, G., 2004, ApJL (submitted),
astro-ph/0409650

\bref
Ratra, B., Peebles, P.J.E., 1988, PhRvD, 37, 3406

\bref
Reiprich, T.H.,,  B\"ohringer, H., 2002, ApJ, 567, 716

\bref
Rhodes, C.S., van de Bruck, C., Brax, Ph., Davis, A.-C., 2003, PhRvD,
68, 3511

\bref
Richstone, D., Loeb, A., Turner, E., 1992, ApJ, 363, 477

\bref
Riess, A.G., Filippenko, A.V., Challis, P., et al., 1998,
AJ, 116, 1009

\bref
Riess, A.G., et al. 2004, ApJ, 607, 665

\bref
Rosati, P., Borgani, S., Norman, C., 2002, ARAA, 40, 539

\bref
Szalay, A.S., et al. 2003, ApJ, 591, 1

\bref
Schuecker, P., B\"ohringer, H., 1998, A\&A, 339, 315

\bref
Schuecker, P., B\"ohringer, H., Arzner, K., Reiprich, T.H., 2001a,
A\&A, 370, 715

\bref
Schuecker, P., B\"ohringer, H., Reiprich, T.H., Feretti, L., 2001b, A\&A, 378, 408

\bref
Schuecker, P., et al. 2001c, A\&A, 368, 86

\bref
Schuecker, P., Guzzo, L., Collins C.A., B\"ohringer, H., 2002, MNRAS,
335, 807

\bref
Schuecker, P., B\"ohringer, H., Collins, C.A., Guzzo, L., 2003a, A\&A,
398, 867

\bref
Schuecker, P., Caldwell, R.R., B\"ohringer, H., Collins, C.A., Guzzo,
L., Weinberg, N.N., 2003b, A\&A, 402, 53

\bref
Schuecker, P., B\"ohringer, H., Voges, W., 2004, A\&A, 420, 425

\bref
Schuecker, P., Finoguenov, A., Miniati, F., B\"ohringer, H., Briel,
U.G., 2004, A\&A, 426, 387

\bref
Seljak, U., et al., 2004, PhRvD (submitted) astro-ph 0407372

\bref
Sereno, M., Longo, G., 2004, MNRAS, 354, 1255

\bref
Sheth, R.K., Tormen, G., 2002, MNRAS, 329, 61

\bref
Spergel, D., et al. 2003, ApJS, 148, 175

\bref
Steigman, G., 2002 as cited in Peebles, P.J.E., Ratra, B., 2003, RvMP,
75, 559

\bref
Strominger, A., Vafa, C., 1996, PhL B, 379, 99

\bref
Susskind, L., 1995, JMP, 36, 6377

\bref
Suwa, T., Habe, A., Yoshikawa, K., Okamoto, T., 2003, ApJ, 588, 7

\bref
Tegmark, M., Zaldarriaga, M., 2002, PhRvD, 66, 103508

\bref
Thomas, S., 2002, PhRvL, 89, 081301

\bref
t`Hooft, G., 1993, in Salamfestschrift: a collection of talks, World
Scientific Series in 20th Century Physics, Vol.\,4, eds. A. Ali,
J. Ellis and S. Randjbar-Daemi, World Scientific, 1993, e-print gr-qc/9310026

\bref
Vauclair, S.C., et al., 2003, 412, 37

\bref
Viana, P.T.P., Liddle, A.R., 1996, MNRAS, 281, 323

\bref
Vikhlinin, A., Markevitch, M., Murray, S.S. 2001, ApJ, 551, 160

\bref
Vikhlinin, A., et al., 2003, ApJ, 590, 15

\bref
Visser, M., 1997, PhRvD, 56, 7578

\bref
Vogeley, M.S., Szalay, A.S., 1996, ApJ, 465, 34

\bref
Voit, G.M., 2004 RMP (in press), astro-ph/0410173

\bref
Wald, R.M., 1984, General Relativity, The University of
Chicago Press, Chicago and London

\bref
Wang, L., Steinhardt, P.J., 1998, ApJ, 508, 483

\bref
Weinberg, S., 1989, RvMP, 61, 1 

\bref
Wetterich, C., 1988, NuPhB, 302, 668

\bref
White, S.D.M., Frenk, C.S., 1991, ApJ, 379, 52

\bref
White, S.D.M., Efstathiou, G., Frank, C.S., 1993, MNRAS, 262, 1023

\bref
White, S.D.M., Navarro, J.F., Evrard, A.E., Frank, C.S., 1993, Natur,
366, 429

\bref
White, S.D.M., Fabian, A.C., 1995, MNRAS, 273, 72

\bref
Witten, E., 2000, hep-ph/0002297

\bref
Wu, X.-P., Chiueh, T., Fang, L.-Z., Xue, Y.-J., 1998, MNRAS, 301, 861

\bref
Zhang, Y.-Y., Finoguenov, A., B\"ohringer, H., Ikebe, Y., Matsushita,
K., Schuecker, P., 2004, A\&A, 413, 49

\bref
Zlatev, I., Wang, L., Steinhardt, P.J., 1999, PhRvL, 82, 896

}

\vfill

\end{document}